\documentclass[preprint2,tighten]{aastex_reep}

\usepackage{epstopdf}
\usepackage{subfigure}  
\usepackage{amsmath}

\usepackage{longtable}

\shorttitle{Heating Durations}
\shortauthors{Reep et al.}

\begin{document}

\title{The Duration of Energy Deposition on Unresolved Flaring Loops in the Solar Corona}

\author[0000-0003-4739-1152]{Jeffrey W. Reep}
\affiliation{National Research Council Postdoctoral Fellow, Space Science Division, Naval Research Laboratory, Washington, DC 20375, USA; \href{mailto:jeffrey.reep.ctr@nrl.navy.mil}{jeffrey.reep.ctr@nrl.navy.mil}}

\author{Vanessa Polito}
\affiliation{Harvard-Smithsonian Center for Astrophysics, 60 Garden Street, MS 58, Cambridge, MA 02138, USA}

\author[0000-0001-6102-6851]{Harry P. Warren}
\affiliation{Space Science Division, Naval Research Laboratory, Washington, DC 20375, USA}

\author{Nicholas A. Crump}
\affiliation{Space Science Division, Naval Research Laboratory, Washington, DC 20375, USA}

\begin{abstract}
Solar flares form and release energy across a large number of magnetic loops.  The global parameters of flares, such as the total energy released, duration, physical size, \textit{etc.}, are routinely measured, and the hydrodynamics of a coronal loop subjected to intense heating have been extensively studied.  It is not clear, however, how many loops comprise a flare, nor how the total energy is partitioned between them.  In this work, we employ a hydrodynamic model to better understand the energy partition by synthesizing \ion{Si}{4} and \ion{Fe}{21} line emission and comparing to observations of these lines with \textit{IRIS}.  We find that the observed temporal evolution of the Doppler shifts holds important information on the heating duration. To demonstrate this we first examine a single loop model, and find that the properties of chromospheric evaporation seen in \ion{Fe}{21} can be reproduced by loops heated for long durations, while persistent red-shifts seen in \ion{Si}{4} cannot be reproduced by any single loop model.  We then examine a multi-threaded model, assuming both a fixed heating duration on all loops, and a distribution of heating durations.  For a fixed heating duration, we find that durations of 100 -- 200\,s do a fair job of reproducing both the red- and blue-shifts, while a distribution of durations, with a median of about 50 -- 100\,s, does a better job.  Finally, we compare our simulations directly to observations of an M-class flare seen by IRIS, and find good agreement between the modeled and observed values given these constraints.
\end{abstract}


\nopagebreak

\section{Introduction}
\label{sec:intro}

At the onset of a solar flare, magnetic reconnection releases pent up magnetic stress, energizing many thousands of tenuous magnetic flux tubes in the corona.  Energy released in the corona is transported to the chromosphere, where a sharp rise in pressure causes plasma to ablate into the corona (``chromospheric evaporation'', \citealt{hirayama1974}), filling these flux tubes with hot, dense plasma, causing the extreme brightenings associated with flares.  Simultaneously, due to conservation of momentum, the increased pressure also causes a down-flow of material deeper into the chromosphere (``chromospheric condensation'', \citealt{ichimoto1984}).  While the process of energy transport in solar flares is well understood \citep{benz2008}, many fundamental questions remain unanswered, particularly concerning the multi-threaded nature of flares.

The first open question is the number of coronal loops which form or over which energy is released during a flare, or whether there is even a well-defined number \citep{magyar2016}.  There is no doubt whatsoever that flares do not release their energy over a monolithic loop, which is obvious from imaging \citep{svestka1982,aschwanden2001,sheeley2004}, from spectral considerations \citep{mcclements1989,mariska1993,hori1998,doschek2005,warren2016}, and from deficiencies in models \citep{reeves2002,warren2006,reep2016}.  It is not trivial to count the number of loops in images of flares, nor define an algorithm to do so, since instruments fundamentally are limited by their spatial resolution, and since there is always a number of loops or other features along the line of sight within any given pixel on an imaging instrument.  Nevertheless, numerous studies have measured the widths of coronal loops (\textit{e.g.} \citealt{antolin2012,brooks2012,brooks2013,winebarger2014,brooks2016}), in order to determine whether the observations are resolved.  Recently, the analysis of \citet{aschwanden2017} found that images taken of loops with the \textit{High-resolution Coronal Imager} (\textit{Hi-C}, \citealt{cirtain2013}) are resolved, while images from instruments such as the \textit{Atmospheric Imaging Assembly} (\textit{AIA}, \citealt{lemen2012}) resolve some but not all loops.  Most of these studies were undertaken in non-flaring active regions, however, which may be fundamentally different in topology from flaring loops.

The next problem concerns the parameters of individual loops within a flaring arcade.  It is implausible that each loop receives an equal amount of energy, but it is unclear what the distribution of energy might be, whether that distribution is stochastic in some way, how the released energy is partitioned amongst various transport mechanisms, \textit{etc}.  Estimates of total flare energies, and how that energy is partitioned, are commonly performed (\textit{e.g.} \citealt{emslie2004,emslie2005,milligan2014}), but these are generally global estimates, or at best, limited to a certain volume of a flare which contains many loops.  For example, measurements of energy flux delivered by an electron beam are routinely performed by the \textit{RHESSI} satellite \citep{lin2002}, which can both give an estimate of ribbon area and the total energy delivered to that ribbon (\textit{e.g.} \citealt{krucker2011}).  Owing to finite spatial resolution (even more limited in HXRs), and to the optically-thin nature of the corona, however, no instrument is currently able to measure the fraction of energy that goes to each loop.

Given the complexity of solar flares, the problem of understanding how energy is released on individual loops may seem insurmountable. Spectroscopic observations, however, hold many clues that can be used to piece together a coherent picture of solar flares.  For example, hydrodynamic simulations of flares show that the transition region and chromospheric down-flows that occur during impulsive energy release must be short lived, between 50 to 75\,s \citep{fisher1987,fisher1989}, as the flows are quickly stopped by the higher density material at lower heights.  In many flares, observations of the \ion{Si}{4} emission lines taken with the \textit{Interface Region Imaging Spectrograph} (\textit{IRIS}, \citealt{depontieu2014}) show down-flows that persist for many hundreds of seconds, in apparent contradiction with the theoretical prediction (see Section \ref{sec:single} for further discussion).  The modeling by \citet{reep2016} indicates that a model with many loops being heated within a single \textit{IRIS} pixel can reproduce this behavior, however.  This suggests that individual flare loops are woefully under-resolved by current instrumentation, and are not likely to be fully resolved in the foreseeable future.

In this paper, we consider the implications of the up-flows observed in high temperature emission lines.  Early flare observations rarely showed the strong up-flows expected from chromospheric evaporation.  Instead, observed line profiles were generally dominated by a stationary component, even during the earliest part of the flare \citep{antonucci1982,mcclements1989,doschek1993,mariska1993,alexander1998}.  With increasing spatial resolution, strong evaporative up-flows have been observed with increasing frequency \citep{czaykowska1999,milligan2009,doschek2013,brosius2013}.  With its high spatial resolution, \textit{IRIS} routinely observes completely blue-shifted profiles of \ion{Fe}{21} \citep{tian2015,polito2015,polito2016,dudik2016,lee2017,li2017,brosius2017}.  As we will see, the magnitude and duration of these high-temperature up-flows are sensitive to the details of the heating.  

We therefore examine various heating durations with hydrodynamic simulations in order to determine what values are consistent or inconsistent with the data.  In Section \ref{sec:single}, we first examine a set of simulations of loops with various energy fluxes and heating durations, finding that heating durations of the order of 300\,s are most consistent with up-flows of \ion{Fe}{21}, though single loops cannot reproduce pervasive red-shifts routinely seen in \ion{Si}{4} emission.  In Section \ref{sec:multi}, we then develop a multi-threaded model to see what values can simultaneously reproduce the emission in both lines.  We find that a large number of threads, with a similar median heating duration and high median energy flux are consistent with published observations.  In Section \ref{sec:obs}, we then present observations of an M-class flare with both \ion{Fe}{21} and \ion{Si}{4} emission (where not saturated), and fit the model to this event, finding consistency with heating durations 50--100\,s and high energy fluxes.  Finally, we discuss the implications of this work in Section \ref{sec:discussion}.

\section{Single Loop Model}
\label{sec:single}

In many observational studies of flares with \textit{IRIS}, red-shifts have been observed in transition region lines (\ion{Si}{4}, \ion{O}{4}, \ion{C}{2}) lasting for ten minutes to over an hour \citep{sadykov2015,brannon2015,li2015,polito2016,sadykov2016,warren2016,zhang2016,li_b_2017}, although not all authors commented on this behavior explicitly.  Previously, studies with the \textit{Coronal Diagnostic Spectrometer} (\textit{CDS}, \citealt{harrison1995}) had detected similar red-shifts in \ion{O}{5} and similar lines \citep{czaykowska1999,brosius2004}, though they were not the primary focus of those works.  The first explanation put forth for these red-shifts is that these are simple events of chromospheric condensation, where the expansion of plasma caused by a rise in pressure from chromospheric heating causes a down-flow of material that balances the momentum of the up-flows \citep{canfield1987,zarro1988}.  This explanation suffices for short-lived red-shifts in H$\alpha$ \citep{ichimoto1984}, \ion{Mg}{2} \citep{graham2015}, \ion{Fe}{2} \citep{kowalski2017}, and other chromospheric lines.  However, it was shown with both hydrodynamic loop modeling and analytic considerations, by \citet{fisher1989}, that condensations can only last for 50--75 seconds, regardless of the strength or duration of heating.  

The apparent contradiction between theory and observations can be resolved if one considers a multi-threaded model, where the observed emission is due to more than one loop \citep{reep2016}.  In that work, it was found that a single loop cannot reproduce the persistent red-shifts seen by the flare reported in \citet{warren2016}, for any value of energy flux, or for repeated heating events on that loop.  A multi-threaded model very naturally reproduces them: each successively heated loop dominates the signal, so that the overlap of many loops within one pixel can show red-shifts lasting longer than one minute.  The simulations were also consistent with the observed emission measure distribution, peak temperatures, and density measurements.  Unfortunately, as that flare was small, there was no detectable emission in \ion{Fe}{21}, which may prove useful as a diagnostic of heating duration.  In that work, it was assumed that each individual thread is heated for only 10 seconds, though there is no compelling reason to choose that duration.  

We now turn our focus towards examining this assumption of heating duration, by comparing Doppler shifts of synthesized spectral lines to observations of two particular lines: \ion{Si}{4} 1402.77\,\AA\ and \ion{Fe}{21} 1354.08\,\AA.  The former line has been used as a diagnostic of condensation (\textit{e.g.} \citealt{warren2016}), while the latter is commonly used as a diagnostic of evaporation \citep{graham2015,polito2015,polito2016}.  We begin with the simplest case: a single loop, with various heating durations and energy fluxes.  Can any combination of energy flux and heating duration reproduce observed trends?

\subsection{Simulation Set-up}
\label{subsec:simulation_setup}

In this work, we have run simulations with the field-aligned HYDrodynamics and RADiation code (HYDRAD, \citealt{bradshaw2003}), which solves the equations of conservation of mass, momentum, and energy along a full loop for a two-fluid plasma constrained to a magnetic flux tube with either uniform or expanding cross-section.  The equations and details of the code can be found in \citet{bradshaw2013}.  The code uses adaptive mesh refinement, which is important for resolving the transition region and areas with large gradients, including shocks in the corona.  The code also solves the equations for non-equilibrium ionization (NEI) states, which can diverge significantly from the equilibrium values with impulsive heating \citep{bradshaw2003b,bradshaw2013b}, and affects both the radiative losses and the ionization states for synthesized spectral lines, which is an important consideration for the work here (all ions in this paper are treated in full NEI).  The chromospheric radiative losses are based on the prescription derived by \citep{carlsson2012}.  In this work, we assume that each loop is vertical relative to the solar surface, semi-circular in shape, for a fixed length of $2L = 60$\,Mm, with a uniform cross-section.  

We assume that the heating is due to a beam of non-thermal electrons depositing their energy in a thick-target plasma via Coulomb collisions, with a functional form following \citet{emslie1978} and \citet{hawley1994}, and implementation details in \citet{reep2013,reep2016b}.  We assume an electron spectrum injected at the apex of the loop (and acting symmetrically on each half of the loop) of the form:
\begin{equation}
\mathfrak{F}(E_{0}, t) = \frac{F_{0}(t)}{E_{c}^{2}}\ (\delta - 2) \times
  \begin{cases}
   0 & \text{if } E_{0} < E_{c} \\
   \Big(\frac{E_{0}}{E_{c}}\Big)^{- \delta}       & \text{if } E_{0} \geq E_{c}
  \end{cases}
\label{sharpdist}
\end{equation}
\noindent where $F_{0}$ is the energy flux carried by the beam (keV\,s$^{-1}$\,cm$^{-2}$), $E_{c}$ the low energy cut-off (keV), $E_{0}$ the initial kinetic energy of an electron (keV), and $\delta$ the spectral index.  We take $E_{c} = 15$\,keV and $\delta = 5$ in this work, though they do vary for different flares (\textit{e.g.} \citealt{sui2007,kontar2008,hannah2008}) and likely from loop to loop, and a model trying to reproduce a specific event would need to use the values appropriate to that event.  We assume that the temporal envelopes of the heating are triangular, with equal rising and falling times.  We have run a total of 341 simulations with durations ranging from 1 to 1000\,s, in increments of 0.1 in log space, and peak energy flux values ranging from $10^{8}$ to $10^{11}$\,erg\,s$^{-1}$\,cm$^{-2}$.

After the simulations have been run, we use a forward model to synthesize spectral lines as might be seen by \textit{IRIS} from the values of density, temperature, bulk velocity, and ionization fractions as a function of position and time along the loop(s).  We follow the methodology of \citet{bradshaw2011}, where the emission is calculated along the loop and then binned according to the size of a pixel along the slit, as if the detector were looking down upon the loop at solar center (no longitudinal effects).  The bulk flow velocities are converted into velocities along the line-of-sight in order to calculate Doppler shifts.  We assume that the emission is optically-thin.  We use the \textit{IRIS} response functions obtained from the SolarSoft functions in IDL.  In the multi-threaded modeling, we assume that the loops are all rooted within the same pixel, and therefore include only foot-point contributions to the emission.  A more general model might include a coronal component as a sort of background contribution.

\subsection{Simulations of Single Loops}
\label{subsec:single_sims}

\begin{figure*}
\begin{minipage}[b]{\linewidth}
\centering
\includegraphics[width=\linewidth]{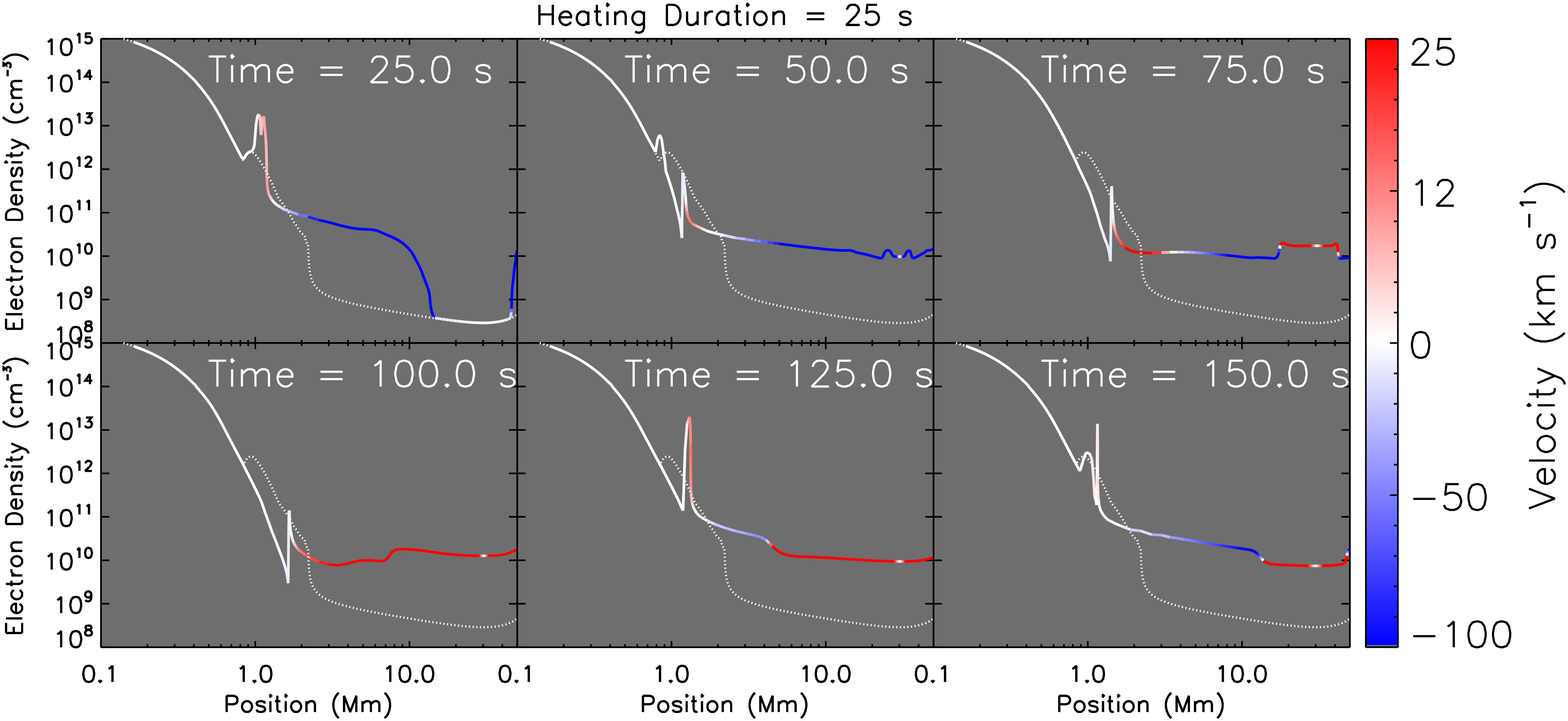}
\end{minipage}
\begin{minipage}[b]{\linewidth}
\centering
\includegraphics[width=\linewidth]{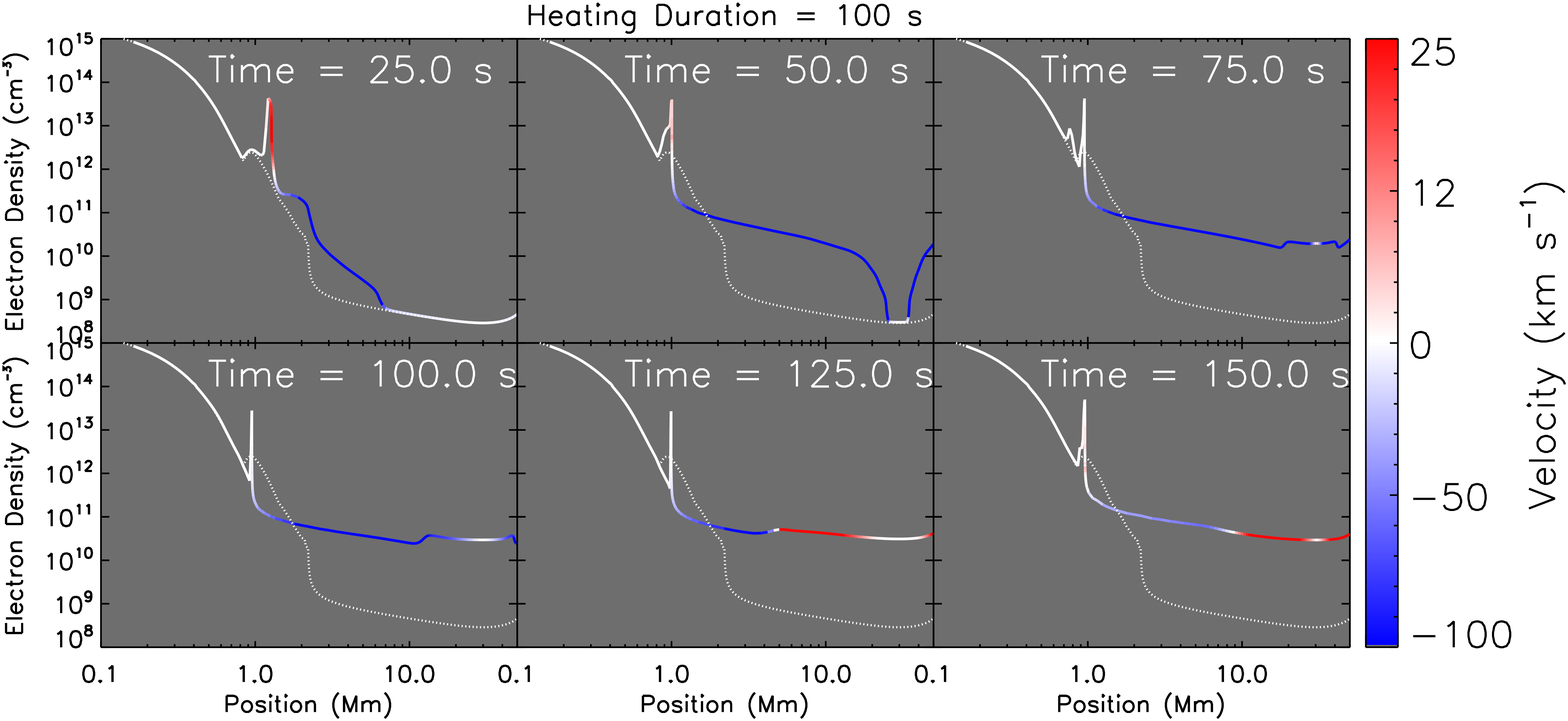}
\end{minipage}
\caption{The electron density as a function of position (logarithmic x-axis), at 6 selected times, for two different simulations with equal energy flux, $F_{0} = 2 \times 10^{10}$\,erg\,s$^{-1}$\,cm$^{-2}$, and heating durations of 25\,s (top) and 100\,s (bottom).  The evolution of the up-flows is strongly dependent on the duration of the heating. The lines are color-coded according to the bulk flow velocity, where red marks down-flows, blue up-flows, and white no significant flows.  The initial profile is shown as a dotted white line for comparison.  Each plot has been truncated at a position just beyond the apex of the loop.}
%
%
\label{fig:density}
\end{figure*}

We begin by showing the hydrodynamic evolution of coronal loops heated by an electron beam for various energy fluxes and heating durations.  In Figure \ref{fig:density} we show a comparison between two simulations heated for the same energy flux, $F_{0} = 2 \times 10^{10}$\,erg\,s$^{-1}$\,cm$^{-2}$, but two different heating durations (\textit{i.e.} two different total energies; 25 seconds on top, 100 seconds on bottom).  Each plot shows the electron density as a function of position (on a logarithmic scale), at the labeled times.  The plots have been colored according to the bulk flow velocity at each position so that red marks the locations where plasma is down-flowing, blue where it is up-flowing, and white where there are no flows.  The dotted white line on each plot shows the initial density profile.  

In the first case with 25 seconds of heating (top plots in Figure \ref{fig:density}), the electron beam quickly deposits its energy into the chromosphere, and the temperature rises sharply.  As the pressure grows, it begins to expand both upwards as an evaporation front, and downwards as a condensation front.  The condensations continue throughout the 25 seconds of heating, only stopping after heating as the temperature begins to fall due to strong radiation in the chromosphere.  The evaporation front, however, flows unimpeded initially, quickly raising the coronal density to well over $10^{10}$\,cm$^{-3}$.  After the heating ceases, the pressure in the chromosphere begins to dissipate, and the evaporation slows to a halt.  After the evaporation fronts from each leg of the loop collide, density waves begin to slosh back and forth across the corona, carrying plasma back down to the chromosphere, and all the while losing energy through the increase in radiative losses.

In the second case (bottom plots in Figure \ref{fig:density}), the heating ramps up more slowly, peaking 50 seconds after the onset.  During that time, the initial condensation front, clearly visible at 25 seconds into the simulation, begins to dissipate as it travels to deeper and denser parts of the chromosphere where the inertia is higher.  Thus, the condensation front dissipates before the heating ceases, unlike the previous case.  The evaporation front continues unimpeded during the period of heating, as in the former case, however.  Because the evaporation lasts longer, the coronal density becomes higher than in the previous case, as well.  After 100 seconds, the heating stops, the evaporation begins to weaken, and the dominant flows are then due to density waves like the previous case. 

\begin{figure*}
\begin{minipage}[t]{0.5\textwidth}
\includegraphics[width=\linewidth]{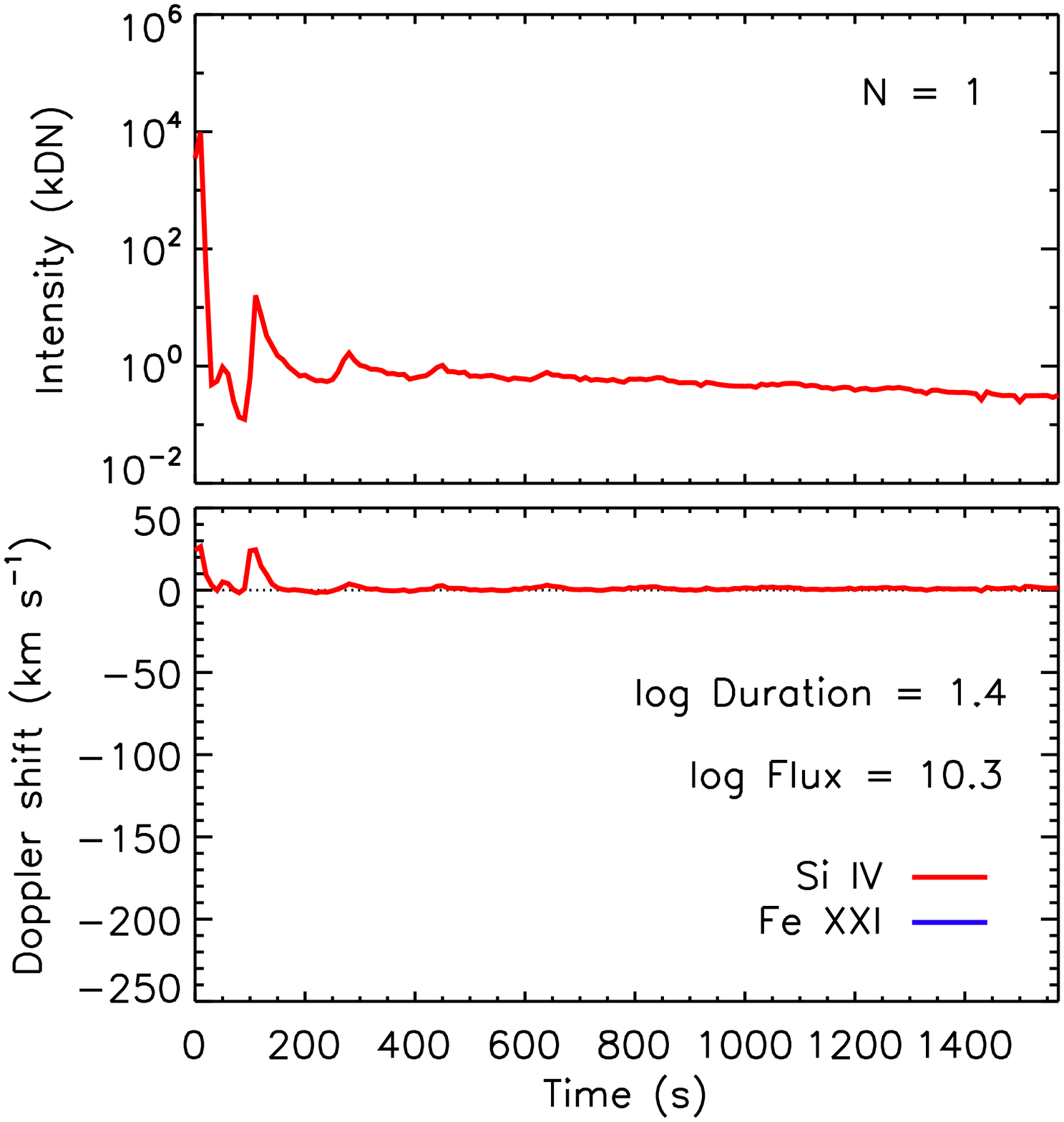}
\end{minipage}
\begin{minipage}[t]{0.5\textwidth}
\includegraphics[width=\linewidth]{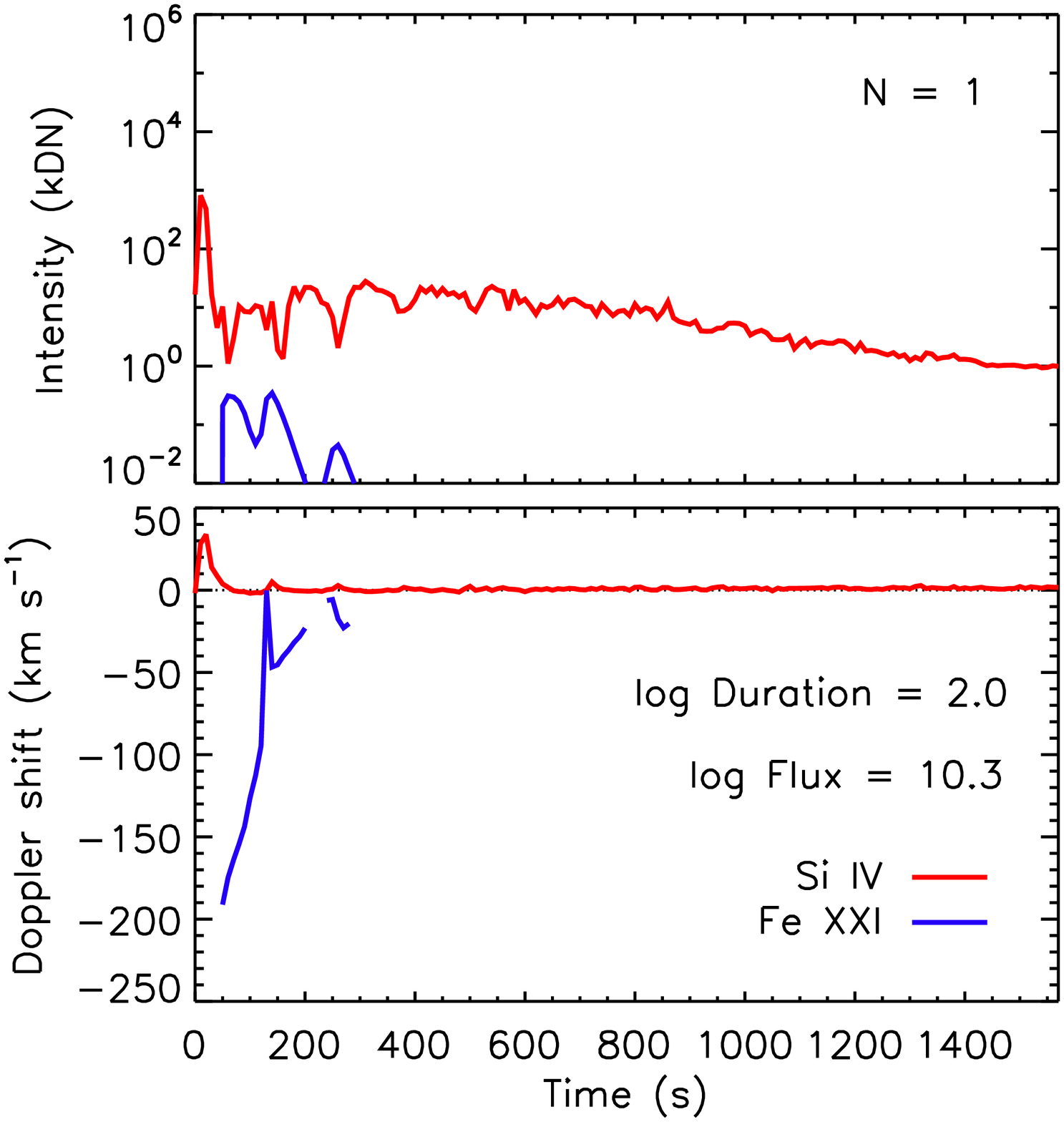}
\end{minipage}
\hspace{\fill}
\caption{The synthesized \ion{Si}{4} (red) and \ion{Fe}{21} (blue) intensities and Doppler shifts near the foot-point of the loop for the simulations in Figure \ref{fig:density}, with the 25\,s case on the left and the 100\,s case on the right.  The note $N = 1$ indicates that these plots are for single loops, rather than multi-threaded simulations.}
\label{fig:single_lines}
\end{figure*}

For these two cases of single loops (number of loops $N = 1$), in Figure \ref{fig:single_lines} we show the synthesized \ion{Si}{4} (red) and \ion{Fe}{21} (blue) line intensity and Doppler shifts along the line-of-sight as a function of time, including the effects of NEI, in the first pixel (\textit{i.e.} near the foot-point, see Figure 1 of \citealt{bradshaw2011}).  The case with 25 seconds of heating is shown at the left, 100 seconds at the right.  In the former, the plasma does not get hot enough to produce \ion{Fe}{21} emission above the noise level that \textit{IRIS} could detect, whereas in the latter there is weak emission that is initially strongly blue-shifted, but quickly slows.  In both cases, the \ion{Si}{4} emission brightens as the plasma is heated, slowly falling off afterwards.  The condensations observed in \ion{Si}{4} quickly decay in both cases.

\begin{figure*}
\begin{minipage}[t]{0.5\textwidth}
\includegraphics[width=\linewidth]{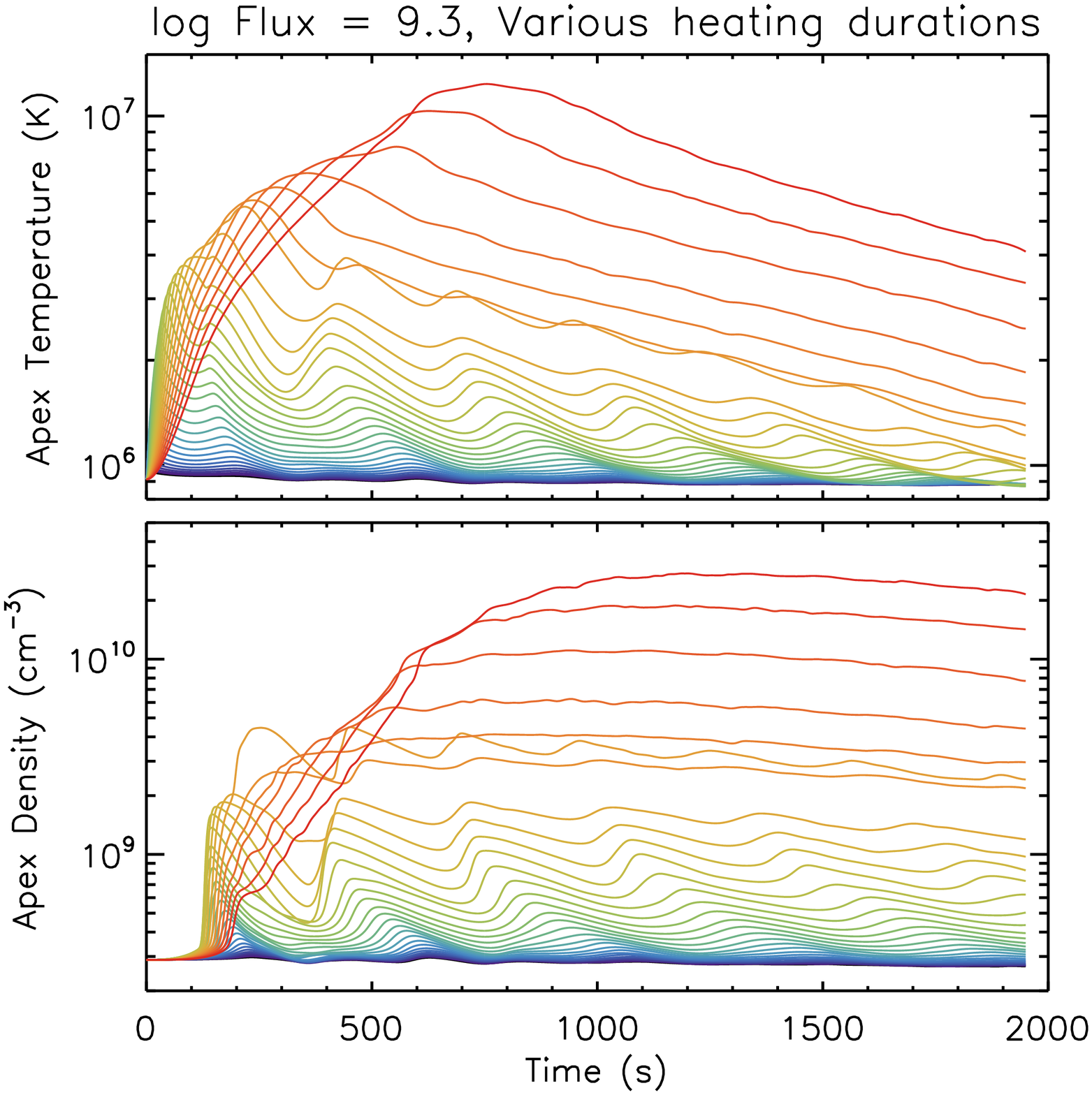}
\end{minipage}
\begin{minipage}[t]{0.5\textwidth}
\includegraphics[width=\linewidth]{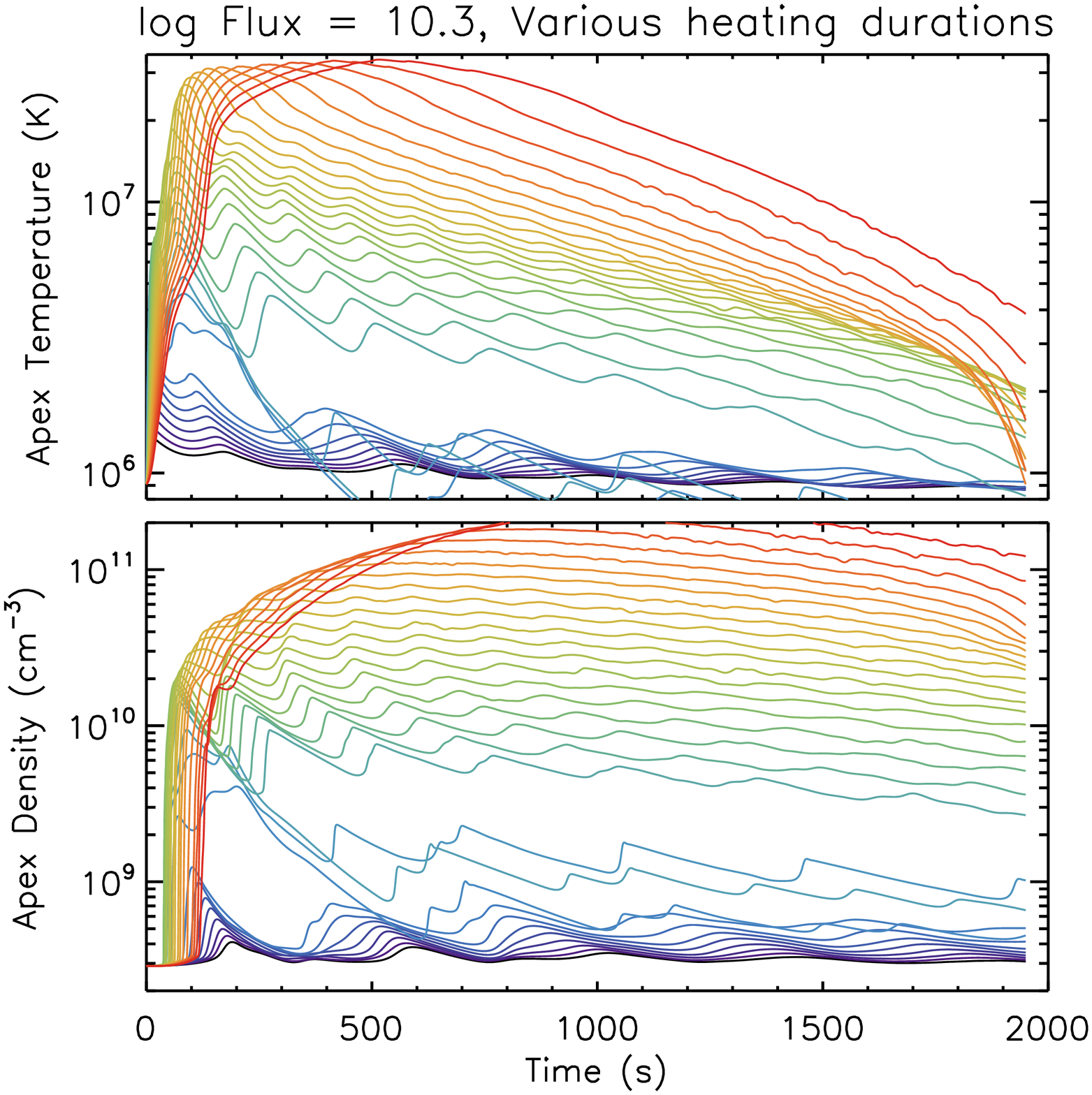}
\end{minipage}
\hspace{\fill}
\caption{The apex temperatures and densities as functions of time in loops heated with an energy flux $\log{F} = 9.3$ (left) and $10.3$ (right) for various heating durations, ranging from 1 to 1000\,s, in increments of 0.1 in log-space (black through red).  The evolution of the plasma depends strongly on both the energy flux and heating duration.}
\label{fig:apex_values}
\end{figure*}

The energy flux and heating duration both strongly affect the plasma evolution on flaring loops.  Figure \ref{fig:apex_values} demonstrates this by showing the evolution of a number of loops.  Each plot displays the apex density and temperature as a function of time in loops heated with a fixed energy flux $\log{F} = 9.3$ (left) and $10.3$ (right) and various heating durations ranging from 1 second (black) up to 1000 seconds (red), with durations shown in colors going across the rainbow.  In order to strongly increase the density near the apex of the loop, there must be a strong evaporation front carrying plasma into the corona from the chromosphere.  The longest heating durations produce the largest densities and temperatures, though the peak times are significantly delayed compared to shorter heating events, some of which barely cause a response in the plasma.  

It is obvious that there is a sharp divide between the left and right plots of Figure \ref{fig:apex_values} due to the difference in energy flux.  In the former, only the longest heating durations produce explosive evaporation, while in the latter, all but the shortest events are explosive.  This suggests that, for a given energy flux and low energy cut-off, the heating duration is also an important variable in determining the threshold for explosive evaporation.  The threshold derived by \citet{fisher1985a,fisher1985b,fisher1985c} assumed a low energy cut-off $E_{c} = 20$\,keV and heating duration of 5\,s, showing that for those values, an energy flux $\gtrsim 10^{10}$\,erg\,s$^{-1}$\,cm$^{-2}$ drives explosive evaporation.  \citet{reep2015} examined the effect of the low energy cut-off on the threshold for explosive evaporation, showing that significantly less energy is required to drive explosive evaporation for cut-offs less than 20\,keV, a result which was first observationally confirmed by \citet{gomory2016}.  One motivation of the present work is to more closely examine the second assumption: how does the heating duration impact evaporation?

\begin{figure*}
\begin{minipage}[t]{0.32\textwidth}
\includegraphics[width=\linewidth]{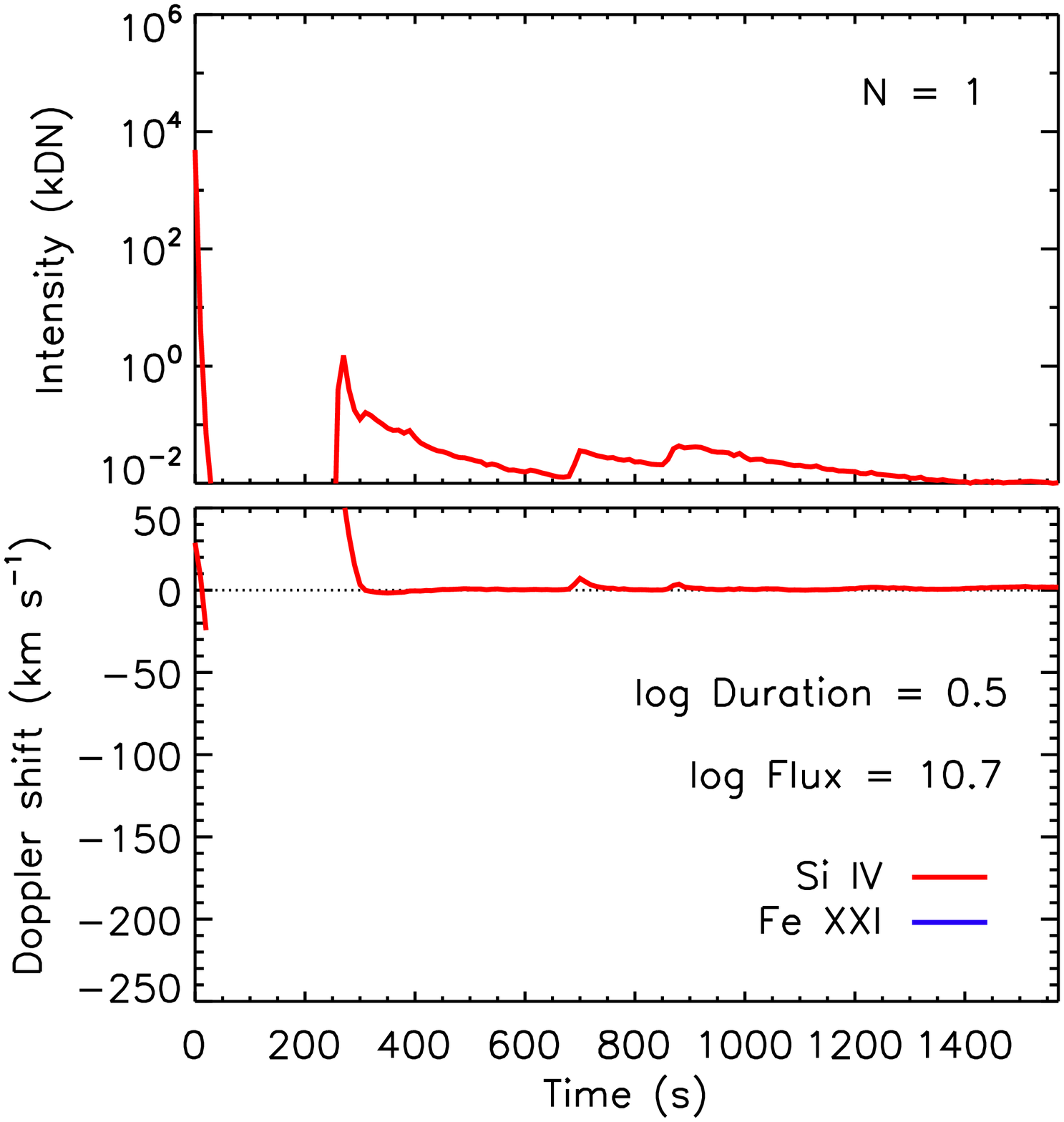}
\end{minipage}
\begin{minipage}[t]{0.32\textwidth}
\includegraphics[width=\linewidth]{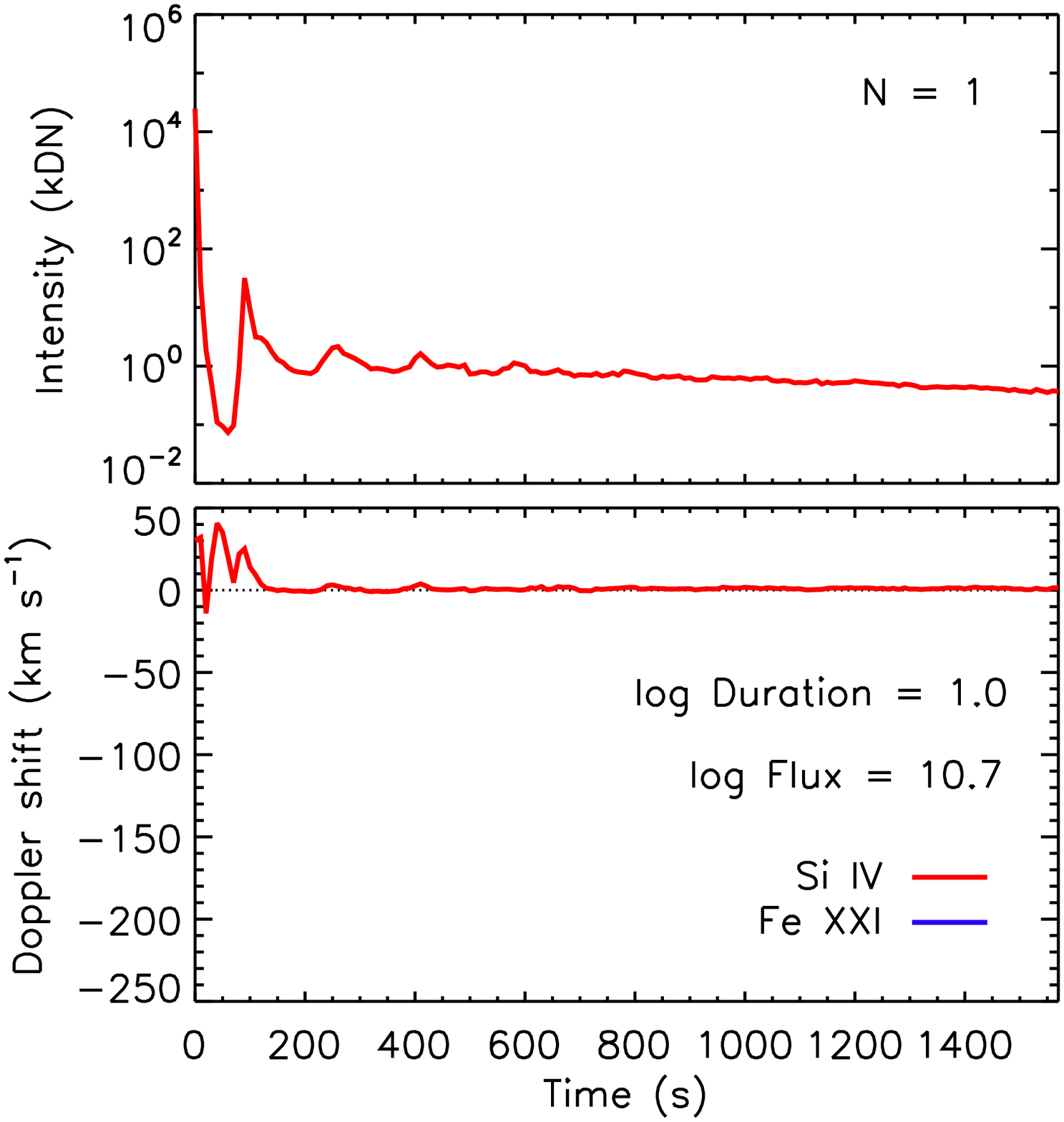}
\end{minipage}
\hspace{\fill}
\begin{minipage}[t]{0.32\textwidth}
\includegraphics[width=\linewidth]{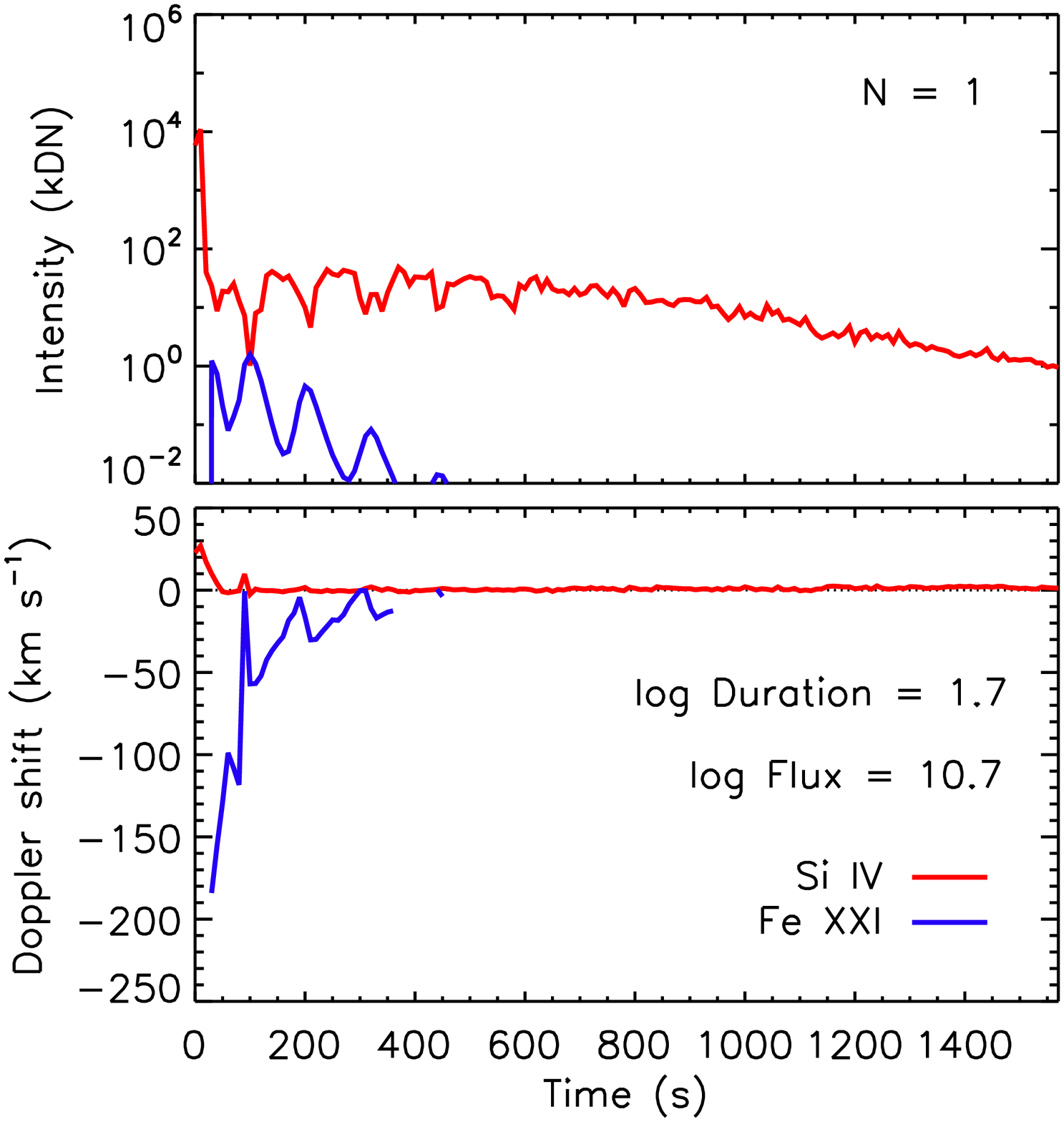}
\end{minipage}
\begin{minipage}[t]{0.32\textwidth}
\includegraphics[width=\linewidth]{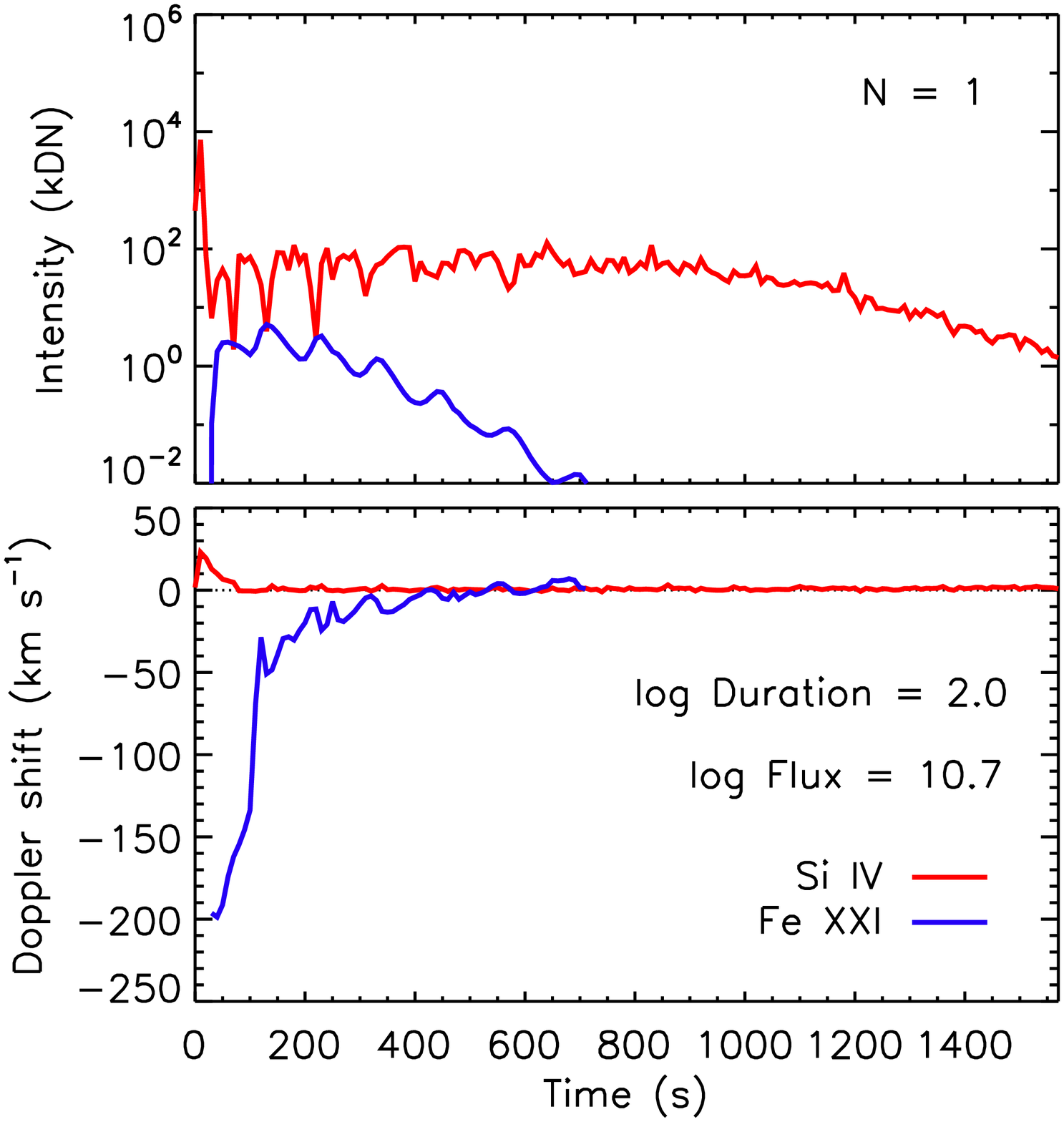}
\end{minipage}
\begin{minipage}[t]{0.32\textwidth}
\includegraphics[width=\linewidth]{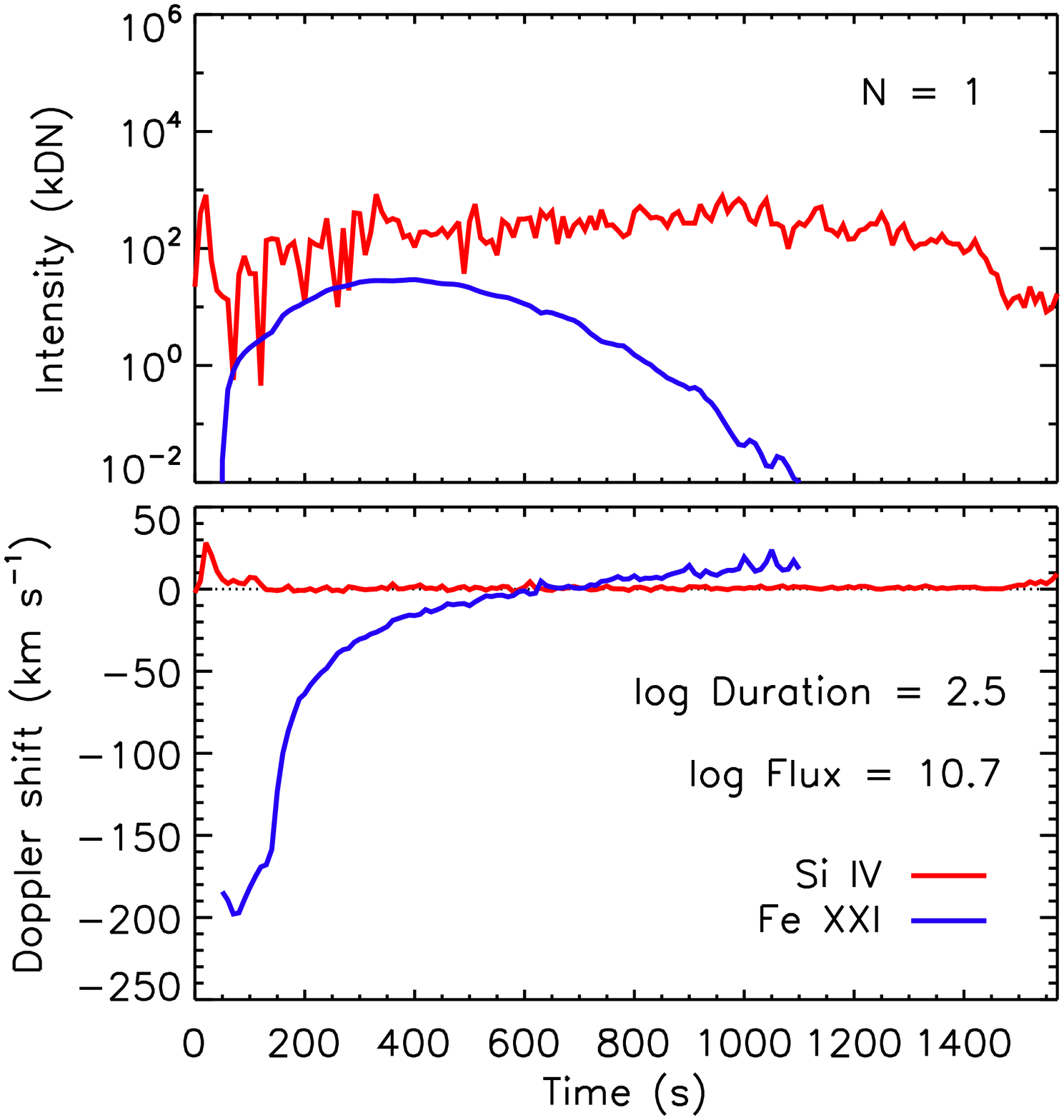}
\end{minipage}
\hspace{\fill}
\begin{minipage}[t]{0.32\textwidth}
\includegraphics[width=\linewidth]{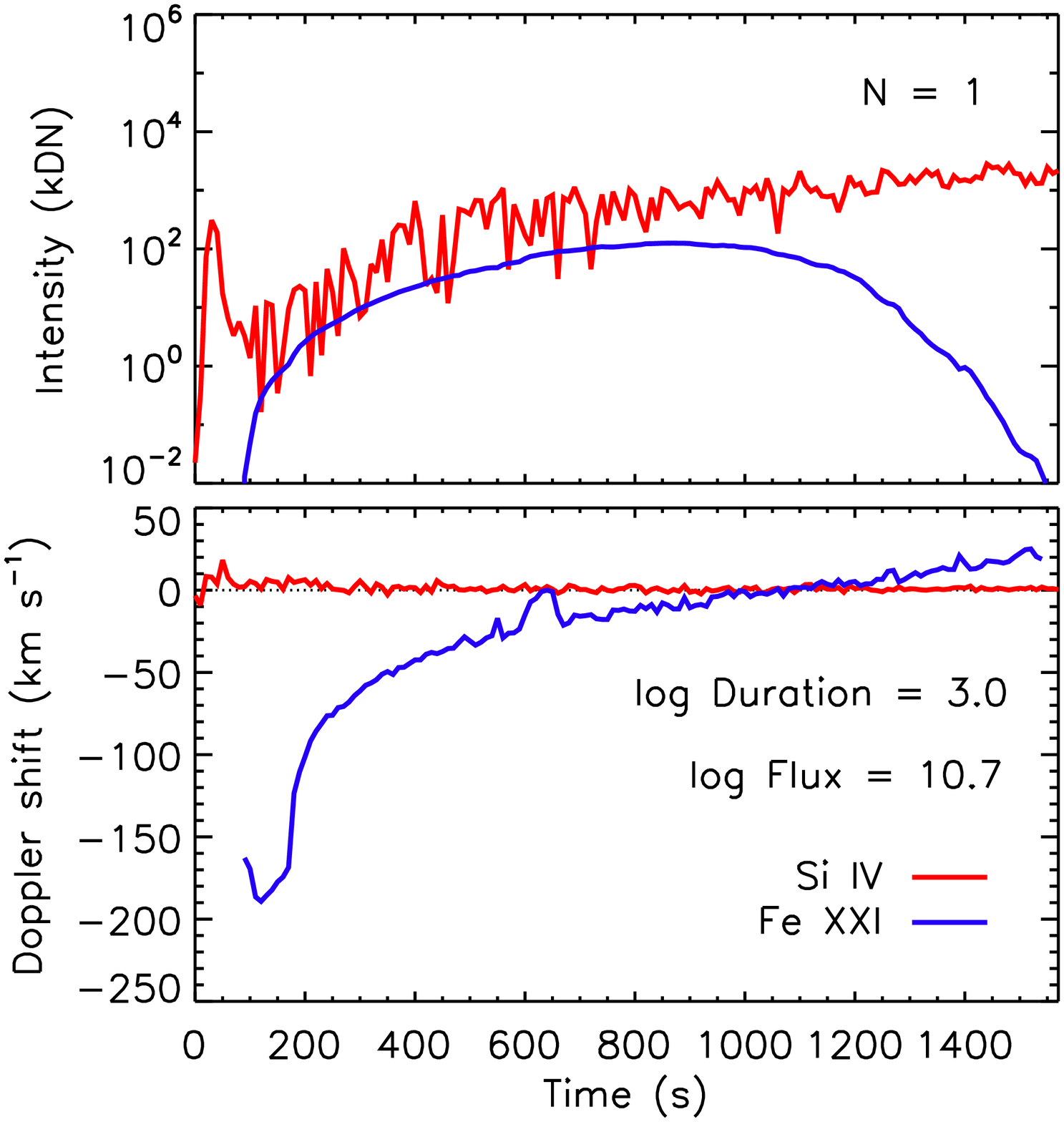}
\end{minipage}
\caption{Synthesized \ion{Si}{4} and \ion{Fe}{21} foot-point emission for 6 simulations with energy flux $F_{0} = 5 \times 10^{10}$\,erg\,s$^{-1}$\,cm$^{-2}$, and heating durations of [3, 10, 50, 100, 316, 1000] seconds (left to right, top to bottom).  The light curves and Doppler shifts of these lines both strongly depend on the heating duration.}
\label{fig:const_flux}
\end{figure*}

We move on to a parameter survey in order to study the effects of heating duration on emission, by first assuming a fixed energy flux and variable heating duration.  In Figure \ref{fig:const_flux}, we show the synthesized foot-point emission of \ion{Si}{4} and \ion{Fe}{21} from 6 simulations with energy flux $F_{0} = 5 \times 10^{10}$\,erg\,s$^{-1}$\,cm$^{-2}$, and heating durations of [3, 10, 50, 100, 316, 1000] seconds (left to right, top to bottom).  

The differences in heating duration have various effects on the emission.  First, in loops heated for a shorter amount of time, there is no visible emission in \ion{Fe}{21}, because the temperature has not increased enough to ionize the iron ions.  This does not imply that there is no evaporation in the loop, only that the evaporation is not visible in the highest temperature lines.  Second, in all cases, the red-shifts in \ion{Si}{4} cease in about one minute, while the blue-shifts in \ion{Fe}{21} take considerably longer to slow.  Further, the time it takes for the blue-shifts to stop is longer for longer heating durations, with an approximately linear correlation.  This can be explained quite readily: once the heating stops, the pressure in the chromosphere begins to fall, so that the expansion which causes evaporation stops.  Third, the intensity of both lines increases with increasing heating duration, simply because more energy has been deposited into the loop.  Fourth, the delay between the onset of heating and the formation of the \ion{Fe}{21} line grows with increasing heating duration, which is due to the assumed temporal envelope with a slow rise to the maximum heating rate.  A constant heating rate for different durations would produce equivalent dynamics while both are still being heated, so that the delay in line formation would be equal.  

\begin{figure*}
\begin{minipage}[t]{0.32\textwidth}
\includegraphics[width=\linewidth]{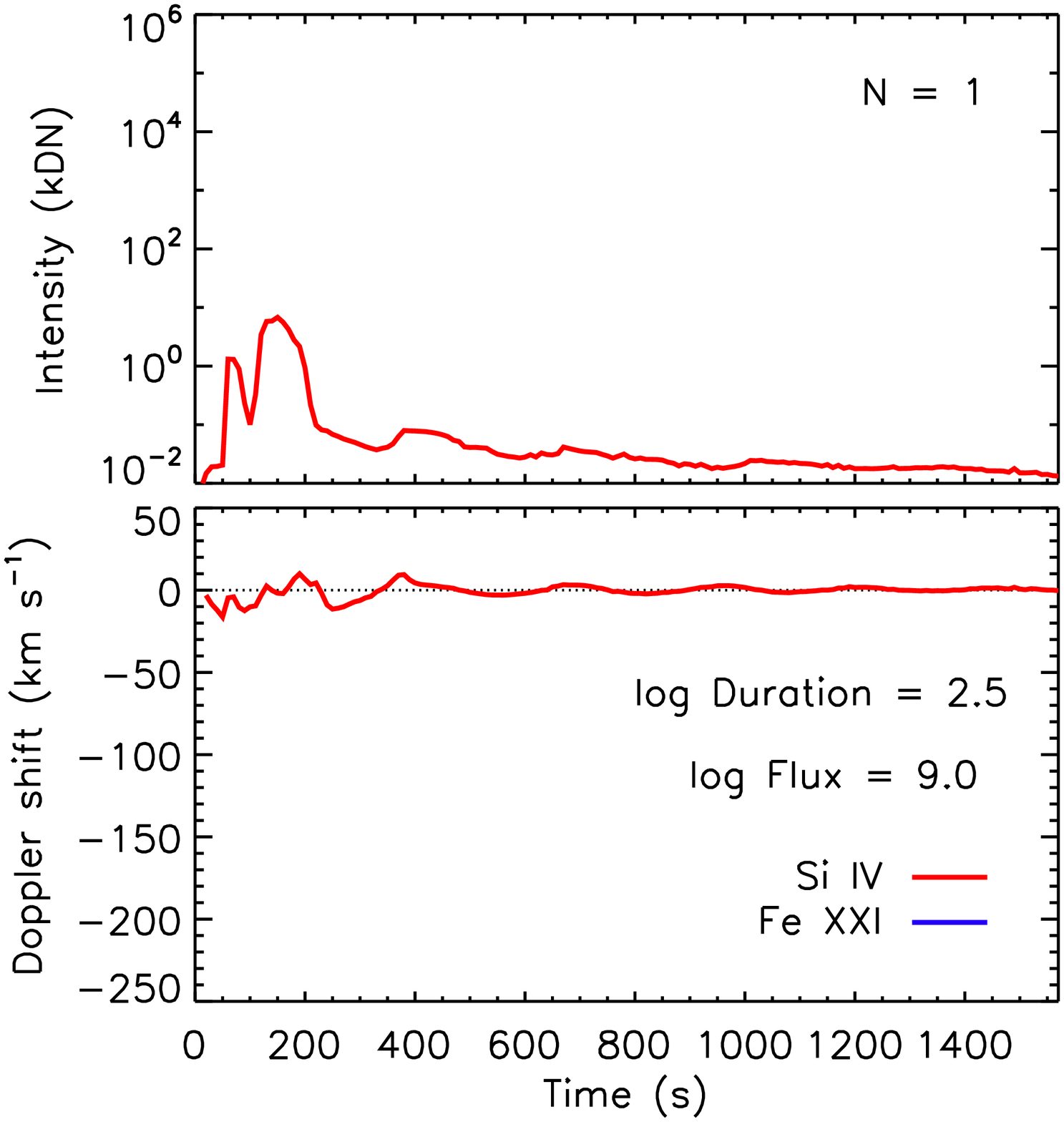}
\end{minipage}
\begin{minipage}[t]{0.32\textwidth}
\includegraphics[width=\linewidth]{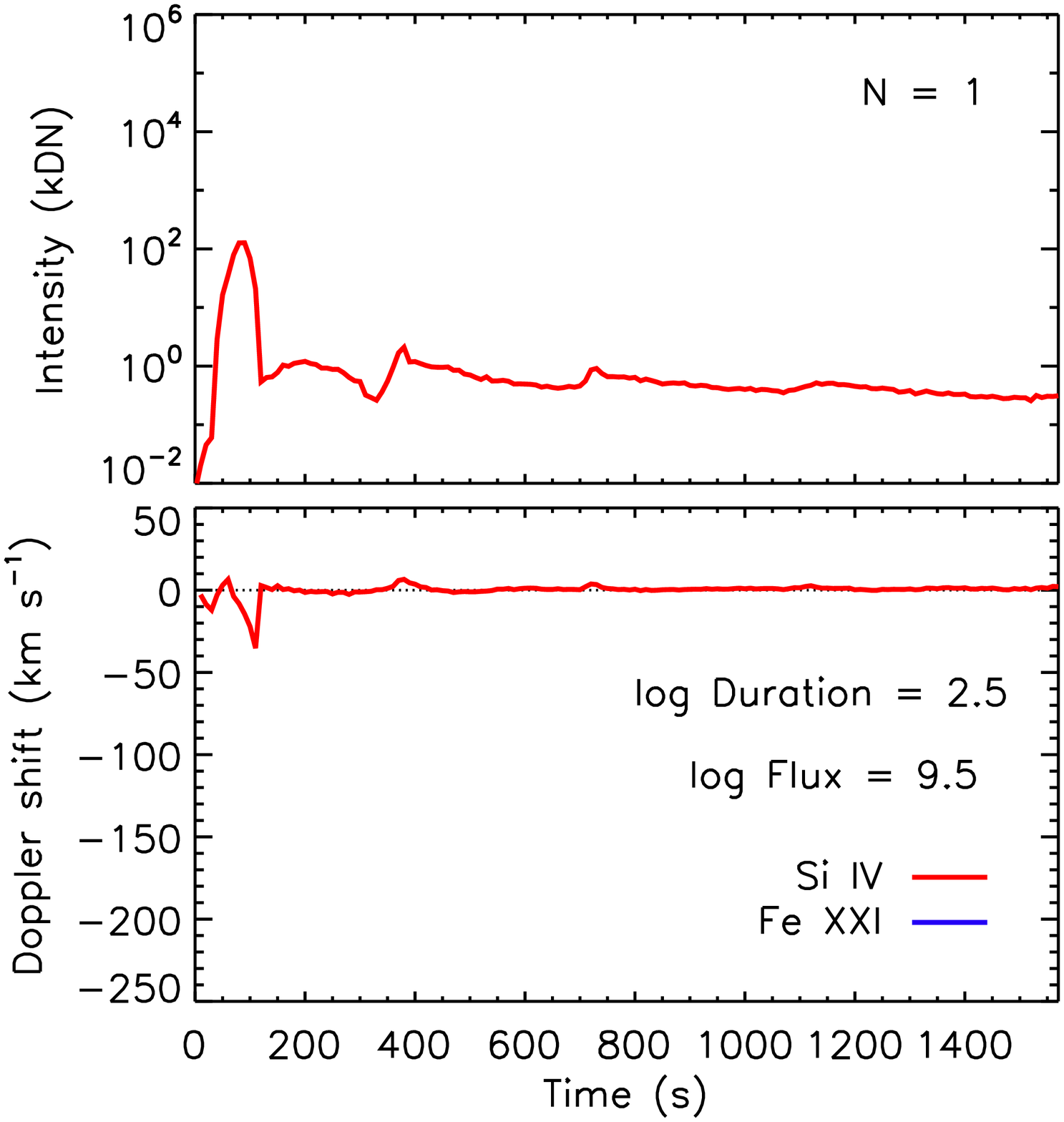}
\end{minipage}
\hspace{\fill}
\begin{minipage}[t]{0.32\textwidth}
\includegraphics[width=\linewidth]{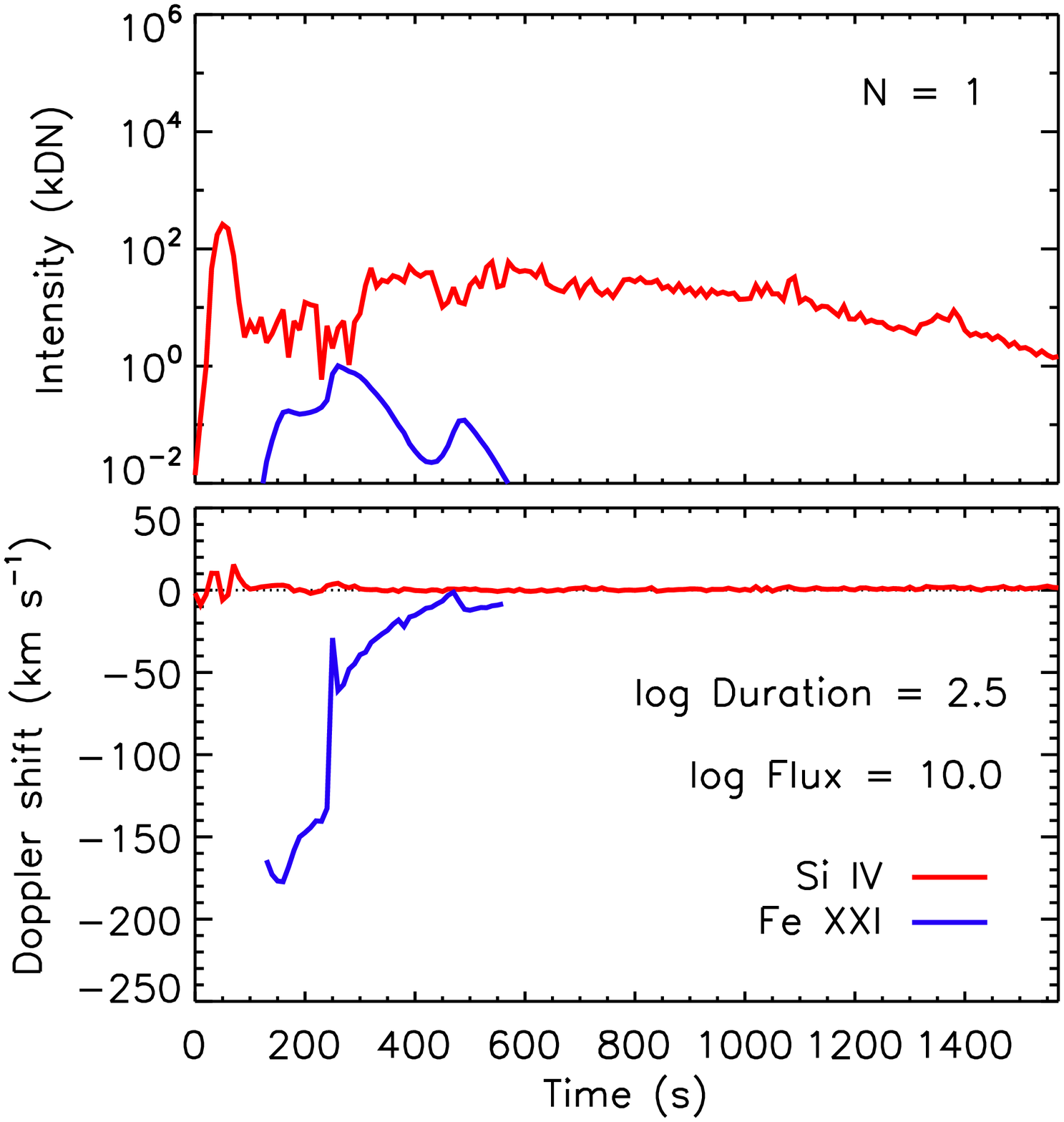}
\end{minipage}
\begin{minipage}[t]{0.32\textwidth}
\includegraphics[width=\linewidth]{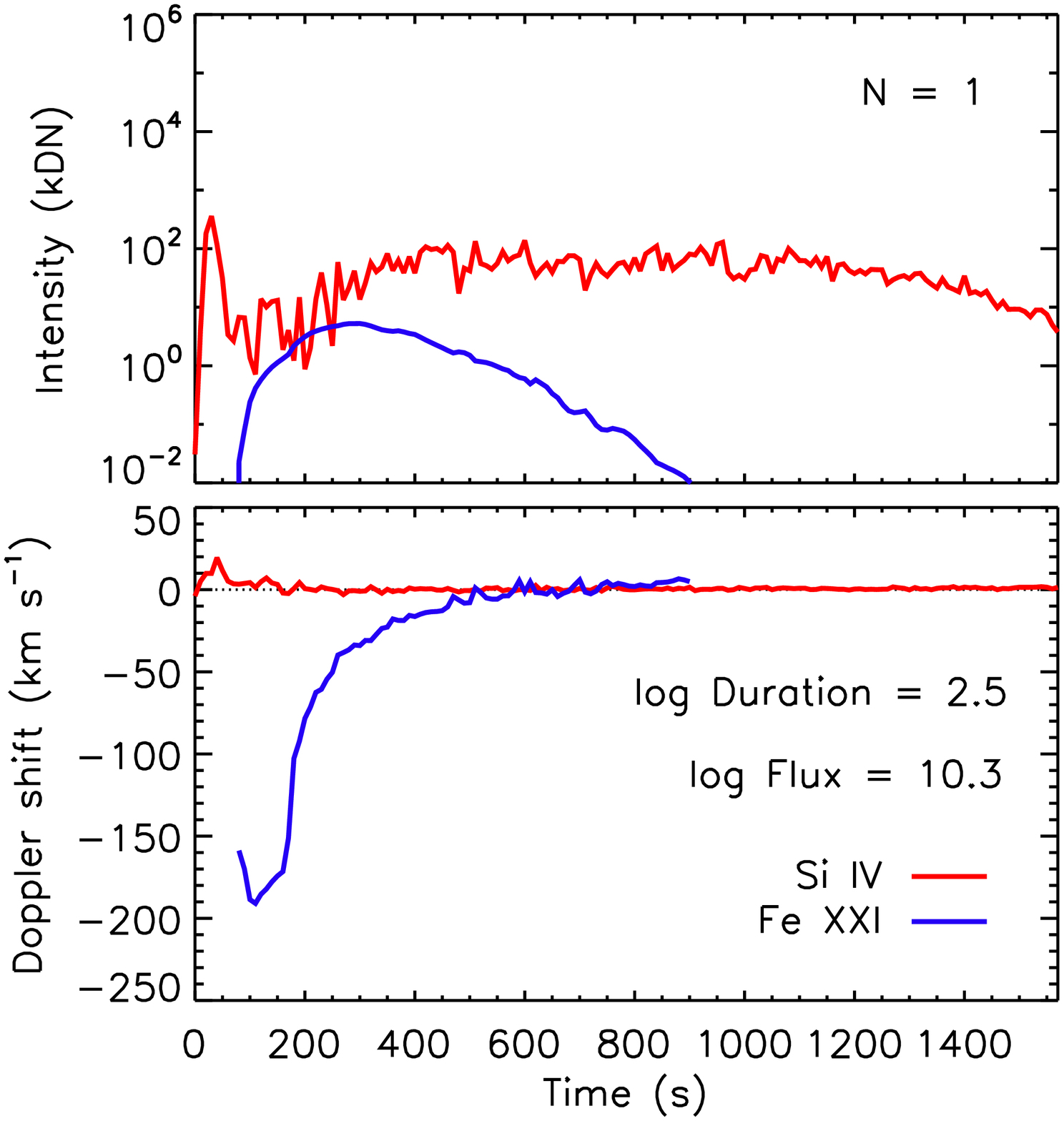}
\end{minipage}
\begin{minipage}[t]{0.32\textwidth}
\includegraphics[width=\linewidth]{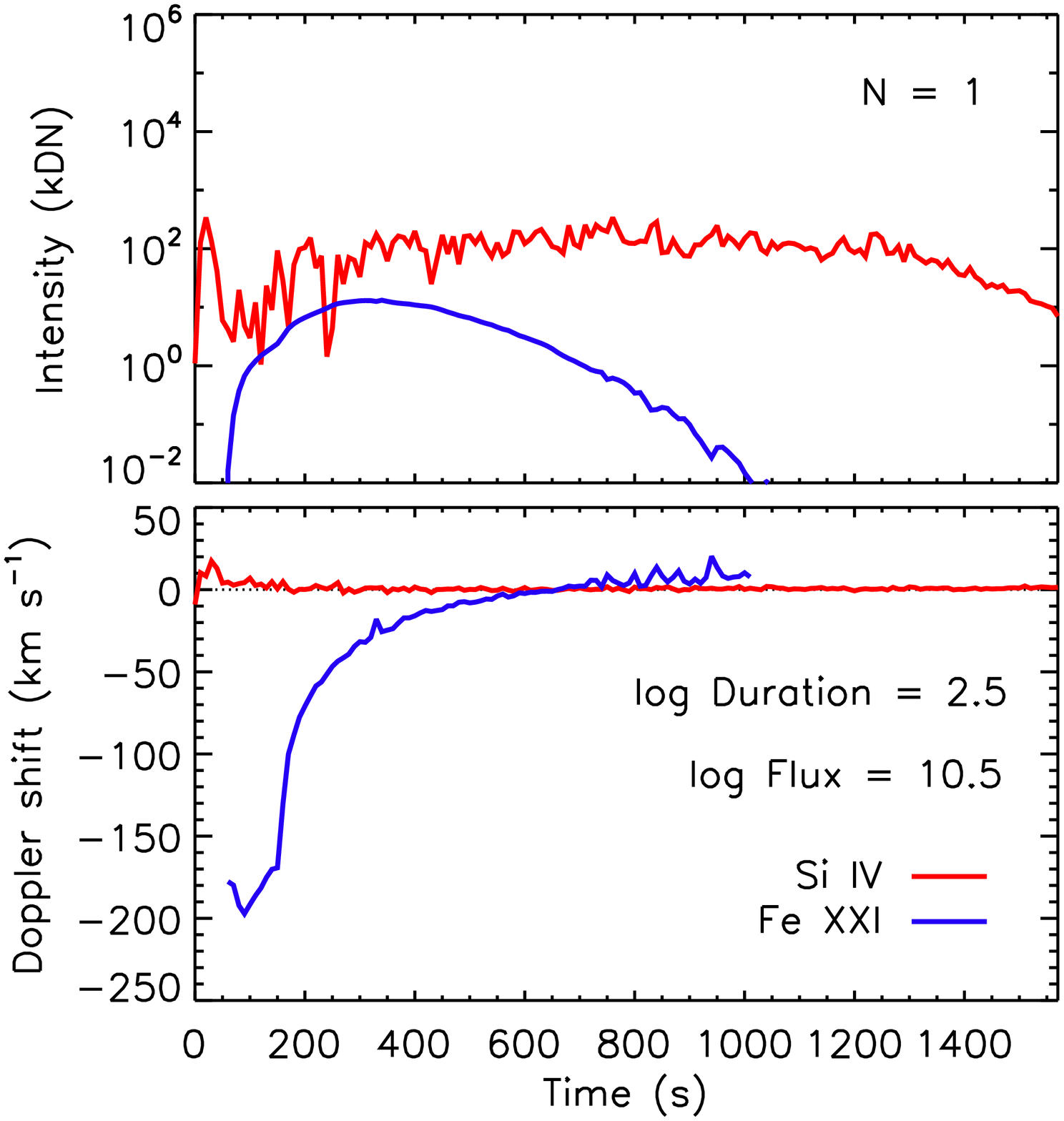}
\end{minipage}
\hspace{\fill}
\begin{minipage}[t]{0.32\textwidth}
\includegraphics[width=\linewidth]{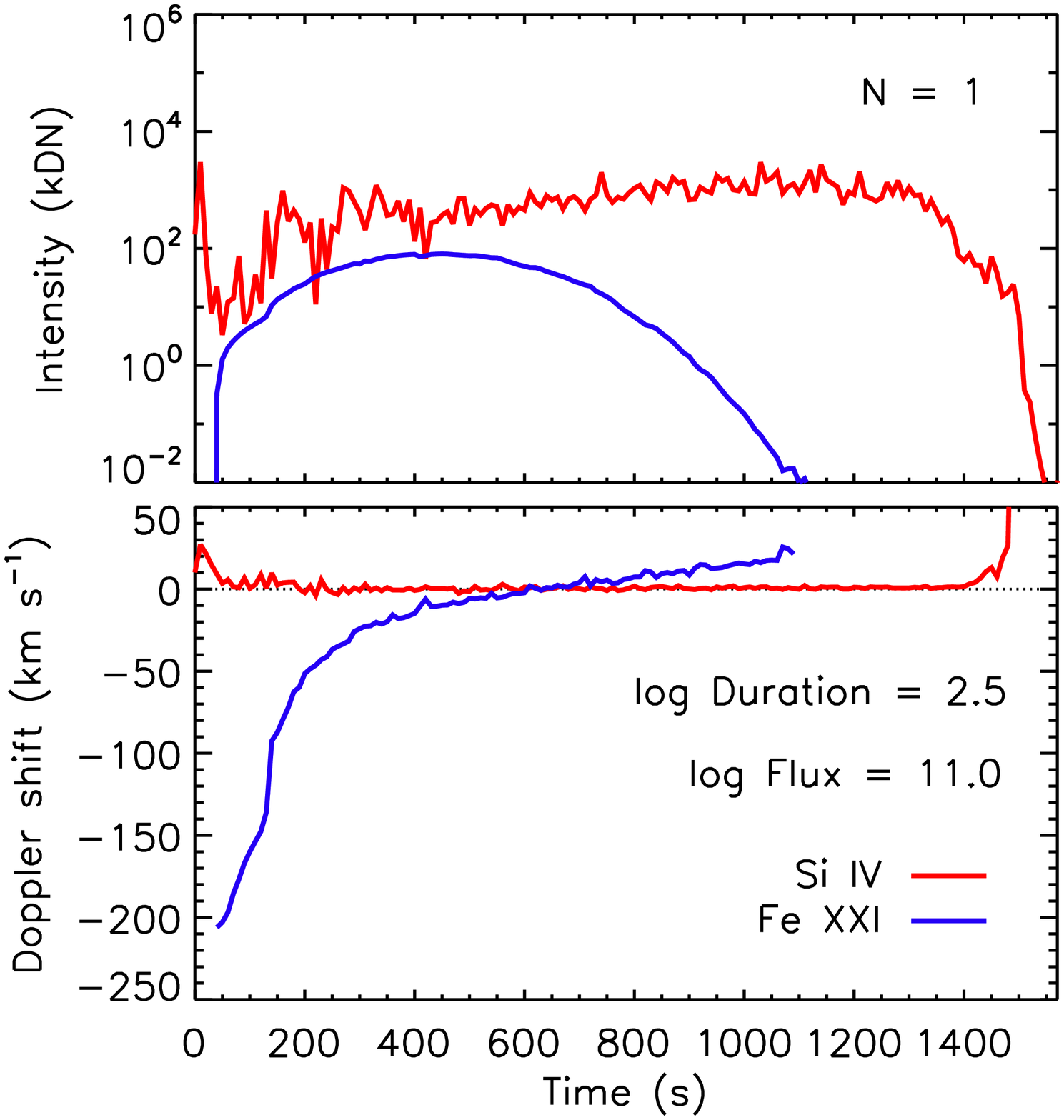}
\end{minipage}
\caption{Synthesized \ion{Si}{4} and \ion{Fe}{21} foot-point emission for 6 simulations with heating duration of 316 seconds, and fluxes $\log{F_{0}} = $[9.0, 9.5, 10.0, 10.3, 10.5, 11.0]\,erg\,s$^{-1}$\,cm$^{-2}$ (left to right, top to bottom).  The light curves and Doppler shifts of these lines both strongly depend on the energy flux.  }
\label{fig:const_dur}
\end{figure*}

We next turn our attention towards studying loops that are all heated for the same duration, 316\,s, but with variable maximum energy fluxes: $\log{F_{0}} = $[9.0, 9.5, 10.0, 10.3, 10.5, 11.0]\,erg\,s$^{-1}$\,cm$^{-2}$.  The synthesized light curves are shown in Figure \ref{fig:const_dur}, from left to right and top to bottom, as before.  

In this case, we again find that there are important trends.  First, for loops that have lower energy fluxes, there is no visible \ion{Fe}{21} emission, as the temperature is again too low.  Second, for lower energy fluxes, \ion{Si}{4} can actually be blue-shifted, rather than red-shifted, which has been reported in IRIS observations \citep{testa2014}, and suggested as a potential diagnostic of the presence of non-thermal electrons in nanoflares.  For higher energy fluxes where the emission is red-shifted, the condensations once again damp in about a minute.  Third, the intensities of both lines increase with increasing energy flux, as one might expect.  Fourth, the delay between the onset of heating and the formation of \ion{Fe}{21} is reduced for larger energy fluxes, as the increased temperature and stronger evaporative flows more quickly ionize the iron ions.  Finally, and perhaps most importantly, in all of the loops where the line forms, the blue-shifts in \ion{Fe}{21} damp in about the same amount of time, suggesting that the heating duration plays a more important role than the energy flux in determining how long evaporation continues.  For this reason, it seems that the heating duration of loops can be diagnosed by the duration of evaporative up-flows.

In all of these cases, however, we see that the condensation flows do not last for more than about one minute.  In flares like the one reported in \citet{polito2016}, where there are both long-lasting red-shifts in \ion{Si}{4} and a slow decrease in the up-flow speed of \ion{Fe}{21}, one explanation could be that a single loop model is insufficient, even though the up-flows alone can be reproduced from a single model with a long heating duration (as done in that paper).  We therefore turn our attention towards a multi-threaded model where we attempt to explain the emission and flows in both lines simultaneously.

\section{Multi-threaded Model}
\label{sec:multi}

We employ a multi-threaded hydrodynamic model in order to understand how the heating on individual loops produces the observed emission in both \ion{Si}{4} and \ion{Fe}{21}.  There are a number of new variables introduced in order to develop a model with more than one loop.  Most importantly, the first is the number of loops $N$ that are rooted within a pixel.  This is essentially a free parameter; we do not know \textit{a priori} what values are reasonable.  Second, there is a small delay in time between the energization of each new loop, which may vary.  We assume that the delay therefore occurs on a Poisson distribution, with an average delay $r$ seconds between events.  In our previous work \citep{reep2016}, we found that the number of loops times that average delay determines the duration of the red-shifts in \ion{Si}{4}, \textit{i.e.} $N \times r \gtrsim \tau_{\text{red-shifts}}$, which led to the conclusion that there were more than 60 loops rooted within an \textit{IRIS} pixel for the event in that study.  

It is improbable that the energy input on each loop is equal, and observations indicate that there is extreme variability from pixel to pixel, which suggests that there is an energy distribution \citep{warren2016}.  The intensities of individual pixels fall on a power-law distribution, so that we assume the energy input similarly is described by a power-law with index $\alpha$.  This index likely varies with time, and from event to event.  In this work, we assume the index is fixed in time, and allow the value to be taken as an input, which is equivalent to modifying the median energy flux of the electron beams injected onto the loops.  

Finally, unlike our previous work where we assumed a fixed heating duration of $10$\,s on all loops, we now allow it to vary.  The event studied in \citet{reep2016} did not produce any \ion{Fe}{21} emission, so that we could not constrain its value.  Because the single loop model indicates that the duration and decay of the evaporative up-flows depend on the heating duration, we use \ion{Fe}{21} emission to constrain heating durations.  We therefore wish to allow the heating durations to vary, and we examine two cases.  In the first case, we assume all loops have the same heating duration.  In the second case, for each loop we randomly select a duration from an assumed distribution with some minimum and maximum values.  The distribution is chosen simply to illustrate the consequences of having variable heating durations, since we do not know whether there is a distribution, or what form it may take.  

We summarize the method by which we create light curves and velocity plots in Table \ref{table:method}.  We wish to emphasize that there are multiple random variables, so the results can change even with the same input parameters, though similar trends are found irrespective of that randomness.  Following this process, we first begin with the case where all loops have the same, fixed heating duration.  We seek to determine whether the multi-threaded model can reproduce both persistent red-shifts in \ion{Si}{4} and the typical up-flow patterns in \ion{Fe}{21}.
\begin{table*}
\caption{The method by which we create the light curves for the multi-threaded model.}
\begin{tabular}{c | l | p{0.5\textwidth}}
Step \# &  Action  &  Description \\
\hline
1 &  Run simulations and forward model &  Create a database of simulations with various values of energy fluxes and heating durations, meant to span the range of reasonable parameters\\
2 &  Choose multi-threaded parameters & Select values: total number of loops $N$, average waiting time $r$, index $\alpha$ of energy power-law distribution, and a minimum energy flux value $F_{\text{min}}$ \\
3 &  Randomly draw energy fluxes  &  Randomly draw $N$ energy fluxes from a power-law distribution with index $\alpha$ and minimum value $F_{\text{min}}$ \\
4 &  Randomly draw heating durations  &  In Section \ref{subsec:fixed_dur}, all heating durations are taken to be equal.  In Section \ref{subsec:distr_dur}, randomly draw $N$ heating durations from an assumed distribution with some minimum and maximum values.  \\
5 &  Randomly draw waiting times & Randomly draw $N$ waiting times from a Poisson distribution, and then offset the start times of each successive loop \\
6 &  Load simulation data & For each pair of energy flux and heating duration, load the appropriate simulation into IDL \\
7 &  Combine spectra & Add the intensities from each loop as a function of time as if they are all located within one pixel, offset each loop by the waiting times, and fit each summed line with a Gaussian at each time to calculate the Doppler velocity \\
8 &  Plot and analyze & \\
\end{tabular}
\label{table:method}
\end{table*}

\subsection{Fixed Heating Duration}
\label{subsec:fixed_dur}

\begin{figure*}
\begin{minipage}[t]{0.32\textwidth}
\includegraphics[width=\linewidth]{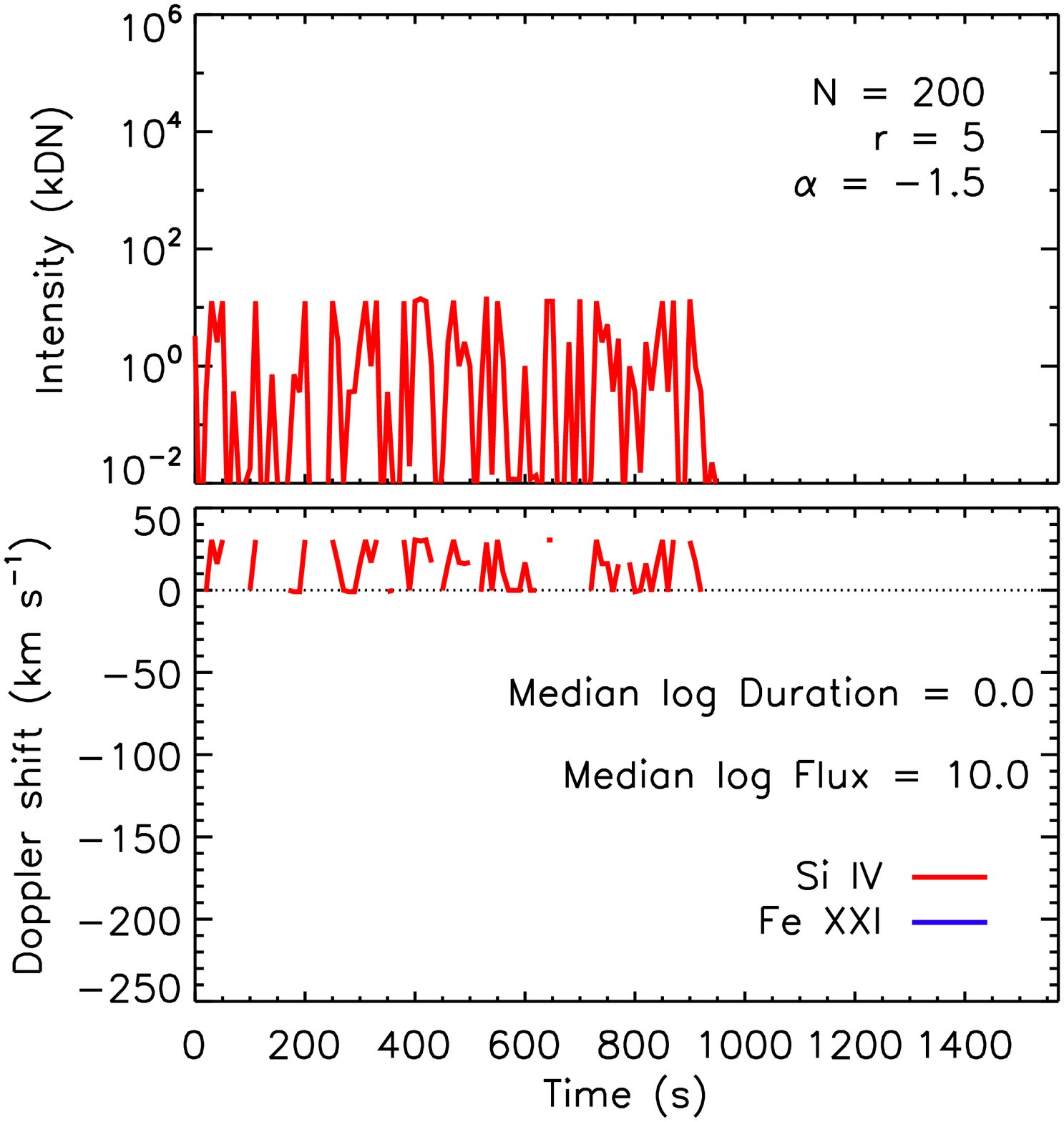}
\end{minipage}
\begin{minipage}[t]{0.32\textwidth}
\includegraphics[width=\linewidth]{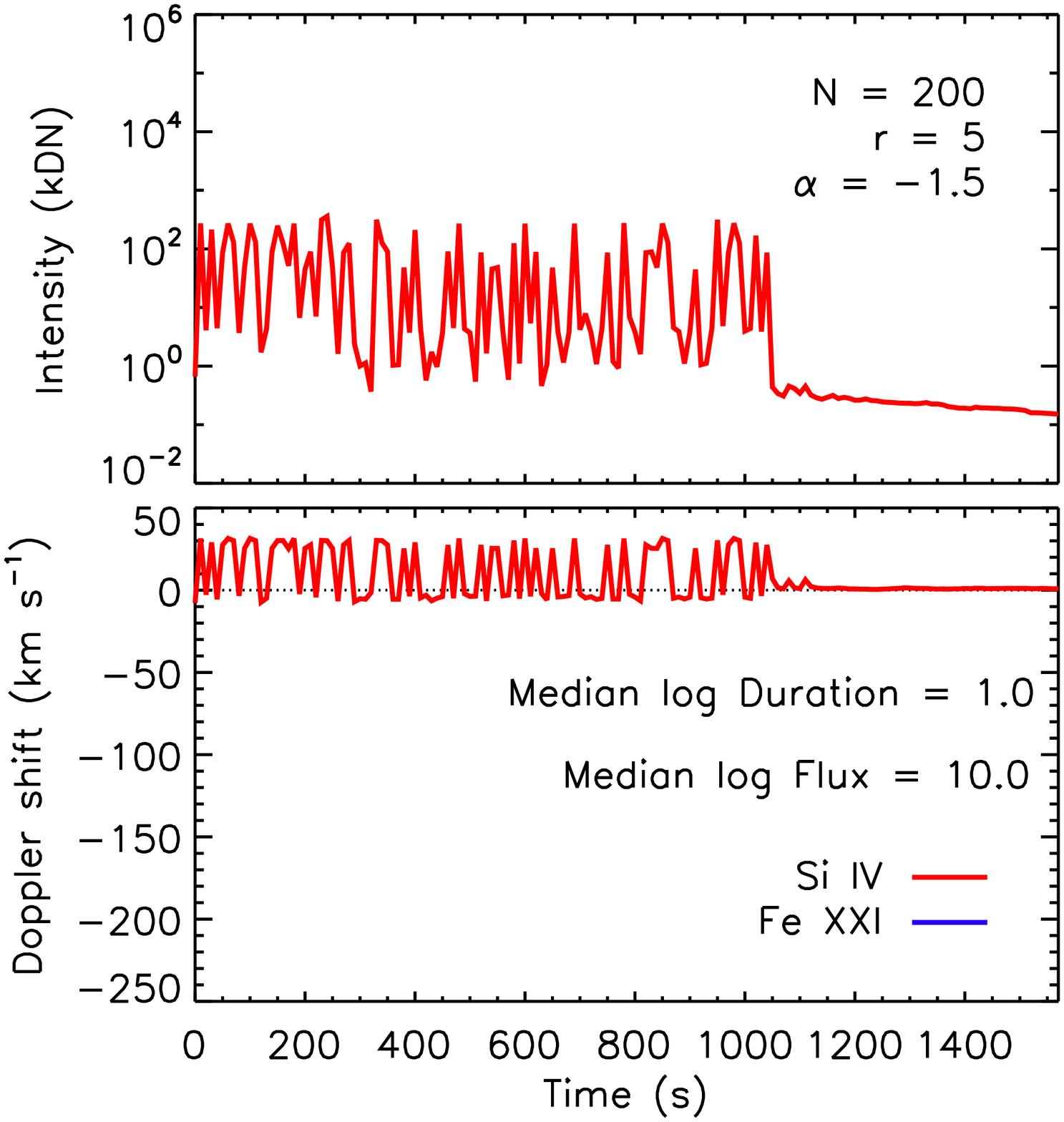}
\end{minipage}
\hspace{\fill}
\begin{minipage}[t]{0.32\textwidth}
\includegraphics[width=\linewidth]{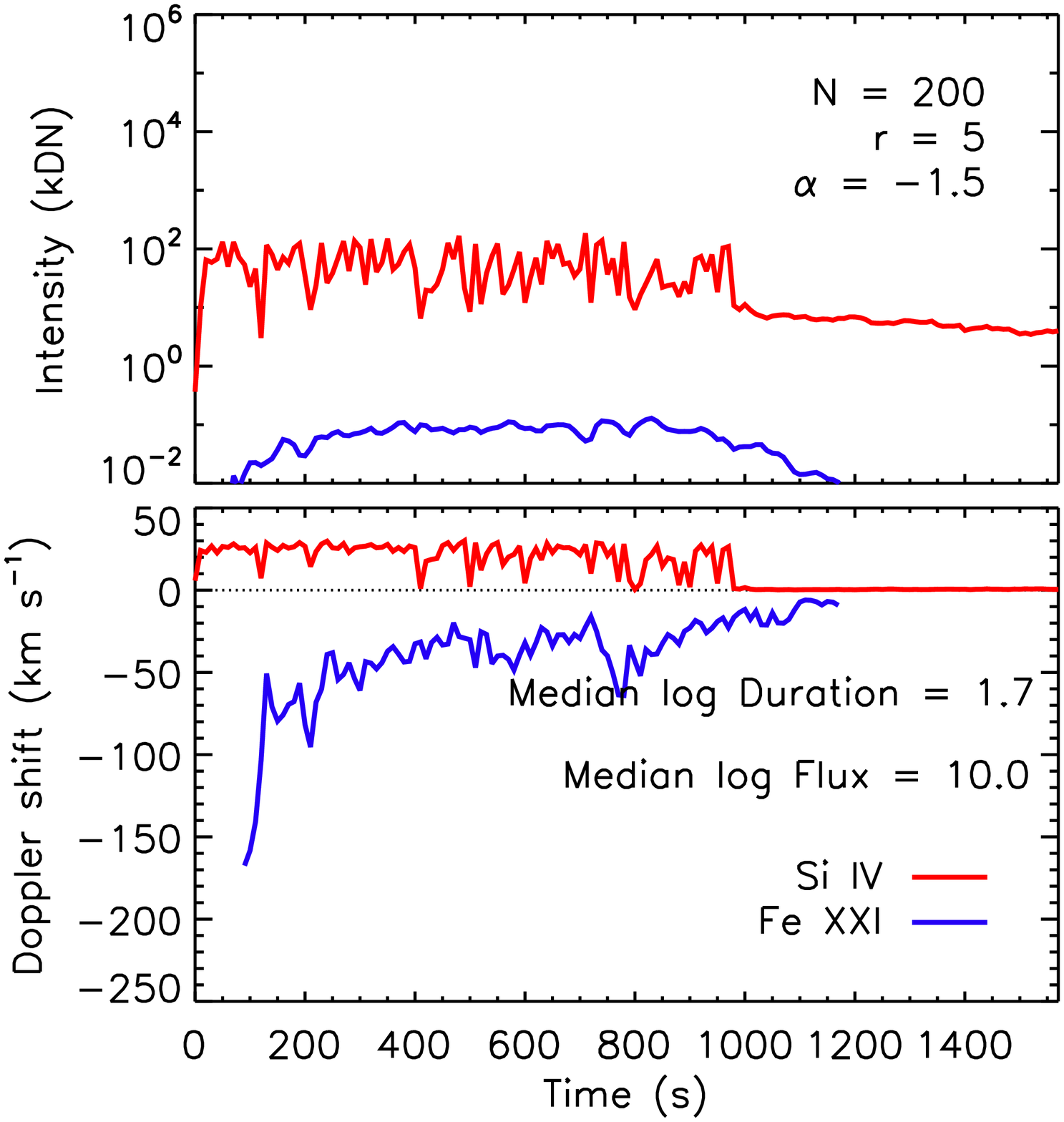}
\end{minipage}
\begin{minipage}[t]{0.32\textwidth}
\includegraphics[width=\linewidth]{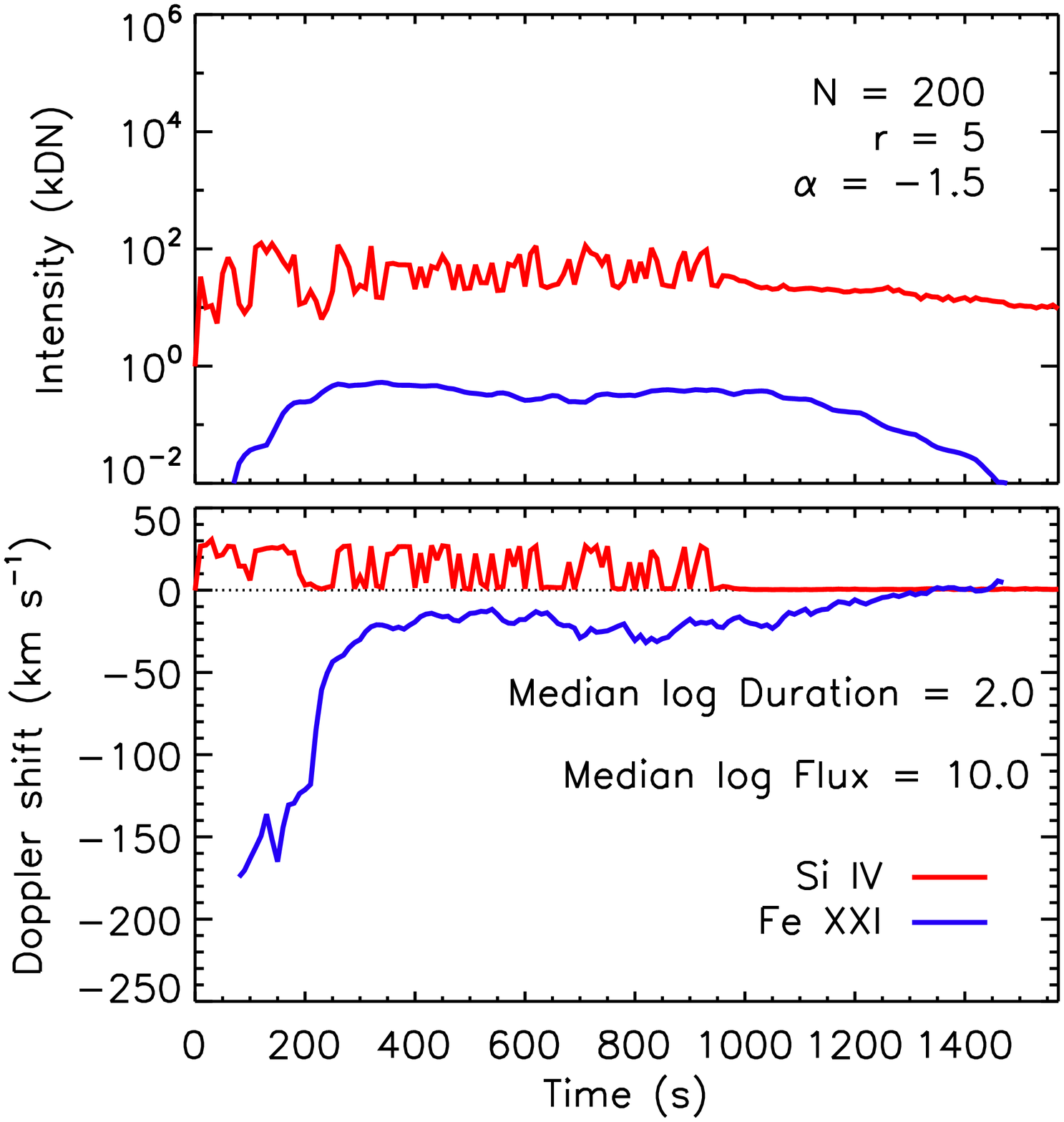}
\end{minipage}
\begin{minipage}[t]{0.32\textwidth}
\includegraphics[width=\linewidth]{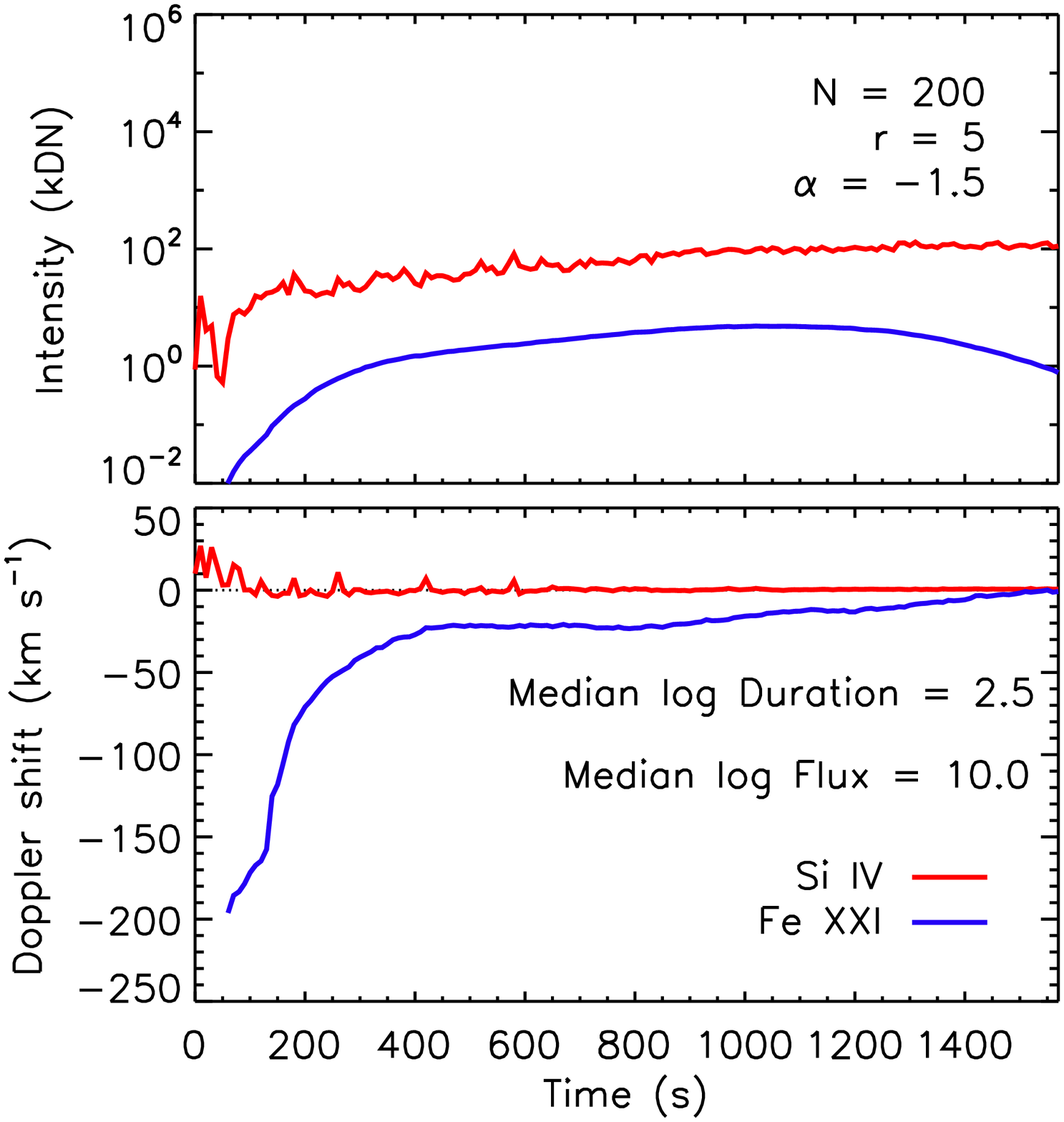}
\end{minipage}
\hspace{\fill}
\begin{minipage}[t]{0.32\textwidth}
\includegraphics[width=\linewidth]{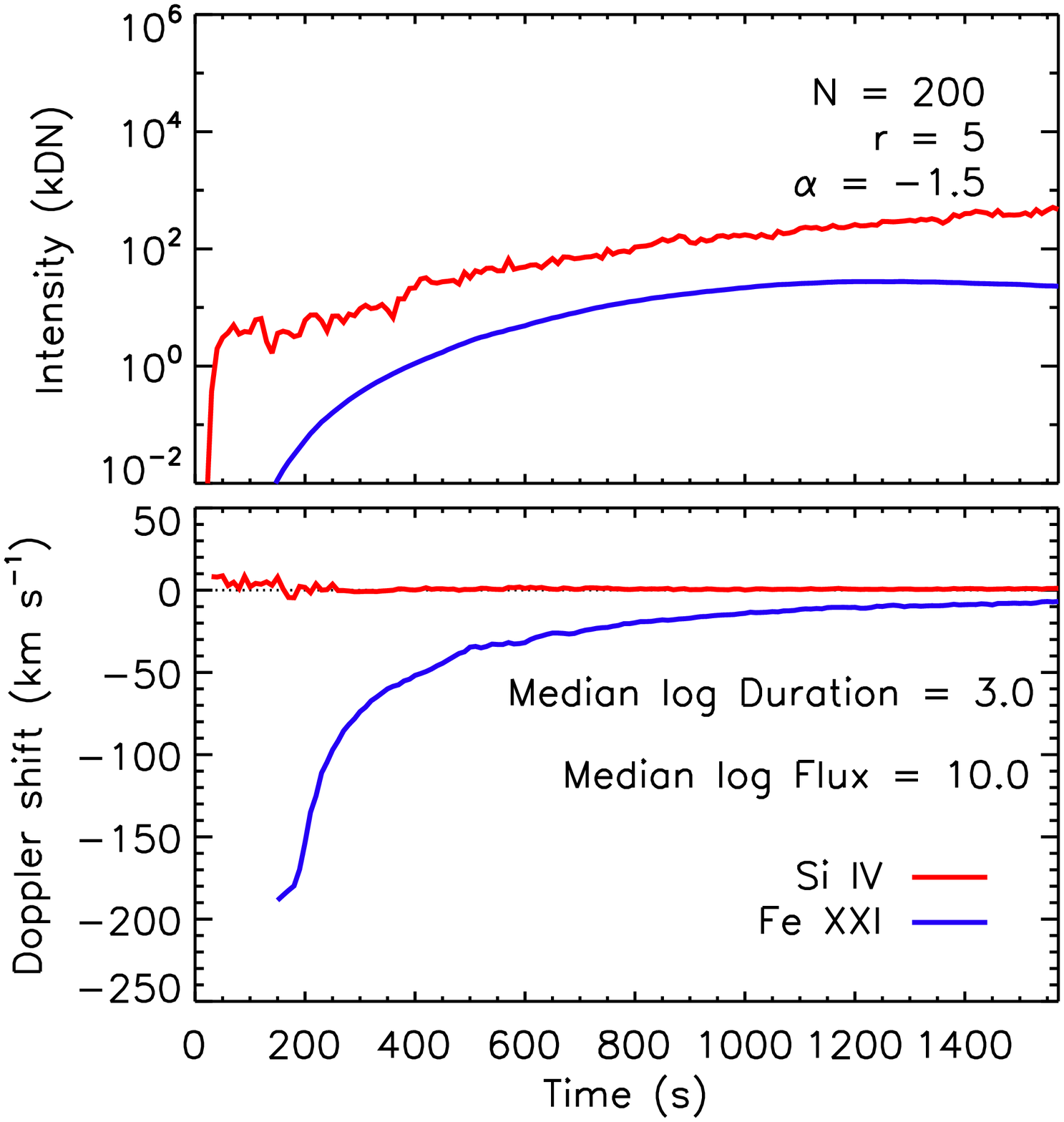}
\end{minipage}
\caption{Synthesized \ion{Si}{4} and \ion{Fe}{21} foot-point emission for 6 multi-threaded simulations with $N = 200$ loops, $r = 5$\,s, $F_{\text{min}} = 3 \times 10^{9}$\,erg\,s$^{-1}$\,cm$^{-2}$, and $\alpha = -1.5$, with heating durations of [1, 10, 50, 100, 316, 1000]\,s (left to right, top to bottom).  Heating durations in the range 50--100\,s appear reasonable compared to observations.  }
\label{fig:N200}
\end{figure*}

We first examine a few cases with equal values of $N$, $r$, $F_{\text{min}}$, and $\alpha$, and various heating durations.  We start with the case where $N = 200$ loops, $r = 5$\,s, $F_{\text{min}} = 3 \times 10^{9}$\,erg\,s$^{-1}$\,cm$^{-2}$, and $\alpha = -1.5$.  Figure \ref{fig:N200} shows results for 6 multi-threaded simulations, where we have set the heating durations to [1, 10, 50, 100, 316, 1000]\,s, from left to right and top to bottom.  

First, consider the \ion{Si}{4} emission.  In all but the shortest case, the intensity peaks slightly above $100$\,kDN, with considerable variability in time.  In the case of short heating durations, the intensity quickly plummets once new loops stop forming, $\approx 1000$\,s after the start.  For longer heating durations, the intensity varies more smoothly, and continues to rise even after the last loop has formed.  In terms of the velocities, shorter heating durations come close to reproducing persistent red-shifts, with the $50$\,s case perhaps the most similar to observations like those of \citet{warren2016}.  For longer heating durations, however, the red-shifts quickly disappear after the start time.  The first few loops to be energized dominate the line signal, and because the heating continues for longer periods of time, the intensity does not fade quickly, though the condensations stop in about a minute, as found in the single loop case.

Next, consider the \ion{Fe}{21} emission.  With the shortest heating durations, the emission is simply not present because the plasma has not been heated to the formation temperature of \ion{Fe}{21}.  For the values assumed, the emission just begins to be visible for heating durations $\gtrsim 20$\,s.  In those cases where it forms, the intensity grows and the light curve becomes considerably smoother with increasing heating duration.  The velocities measured all show the evaporation beginning at around 180--200\,km\,s$^{-1}$, and tending towards 0 at later times.  The rate at which they tend to 0, however, clearly changes, and shorter heating durations more quickly drop in speed.  The cases with longer durations look similar to the single loop case, because the first few loops dominate the signal as they continue to be heated.  

\begin{figure*}
\begin{minipage}[t]{0.32\textwidth}
\includegraphics[width=\linewidth]{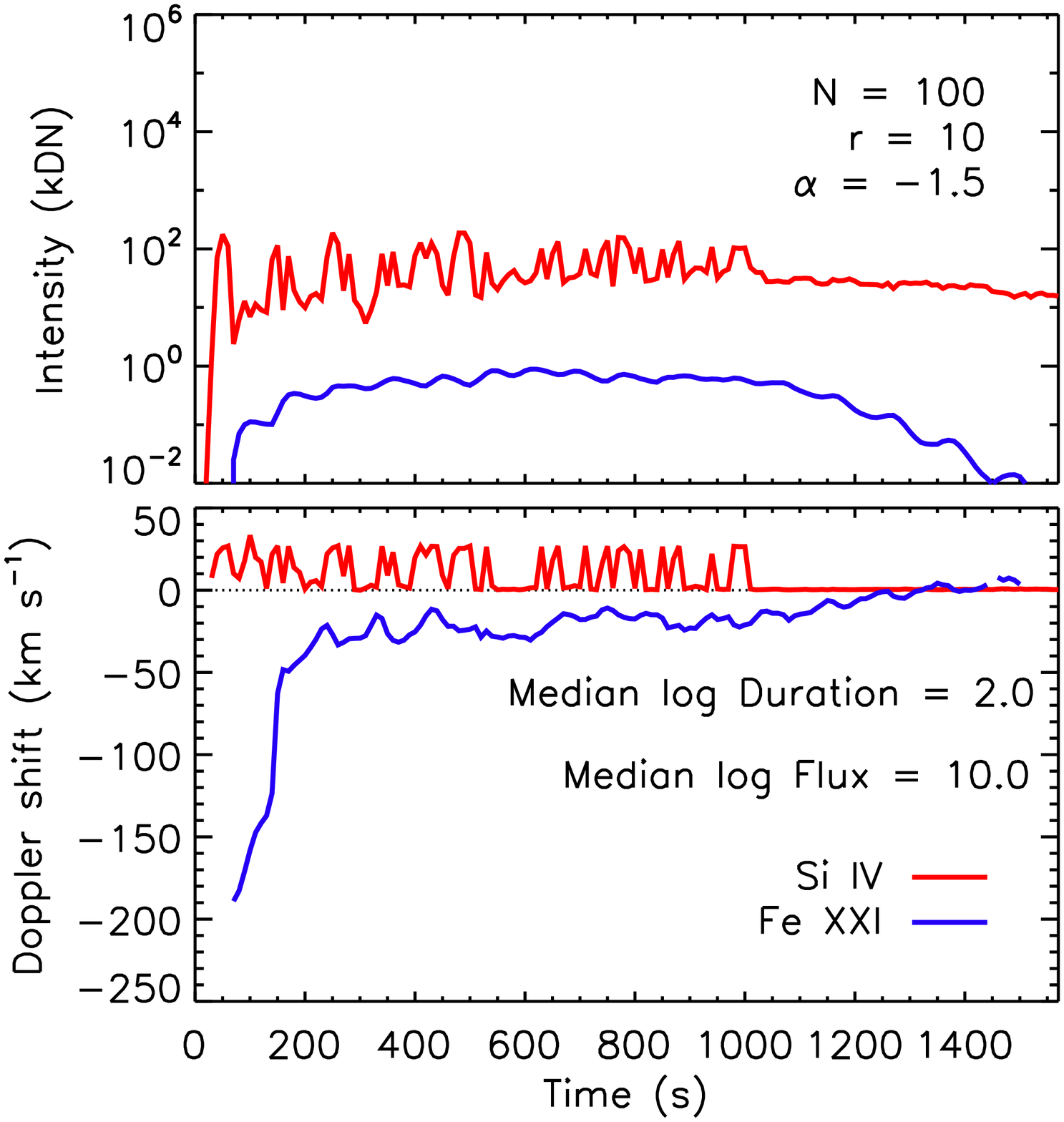}
\end{minipage}
\begin{minipage}[t]{0.32\textwidth}
\includegraphics[width=\linewidth]{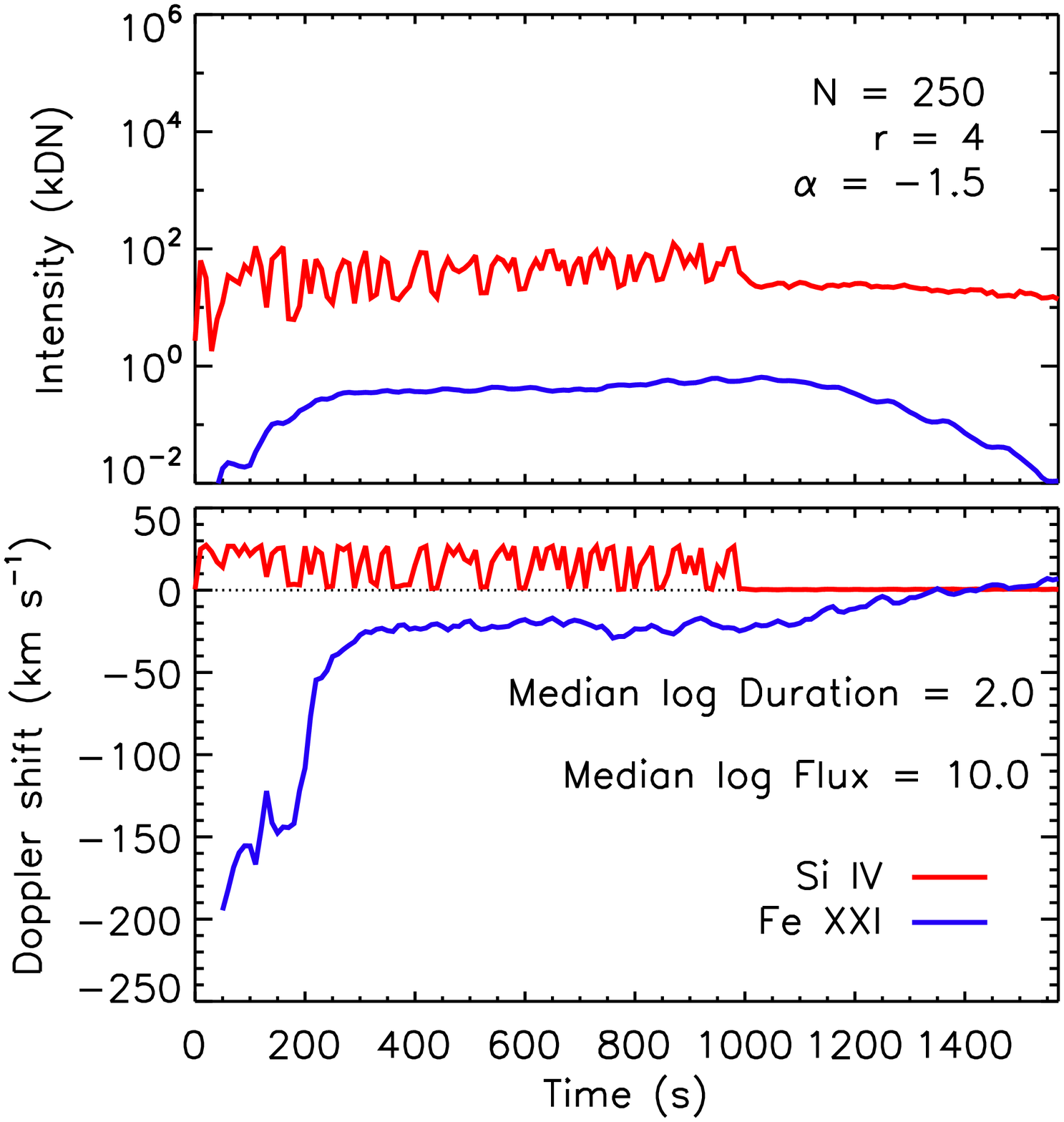}
\end{minipage}
\hspace{\fill}
\begin{minipage}[t]{0.32\textwidth}
\includegraphics[width=\linewidth]{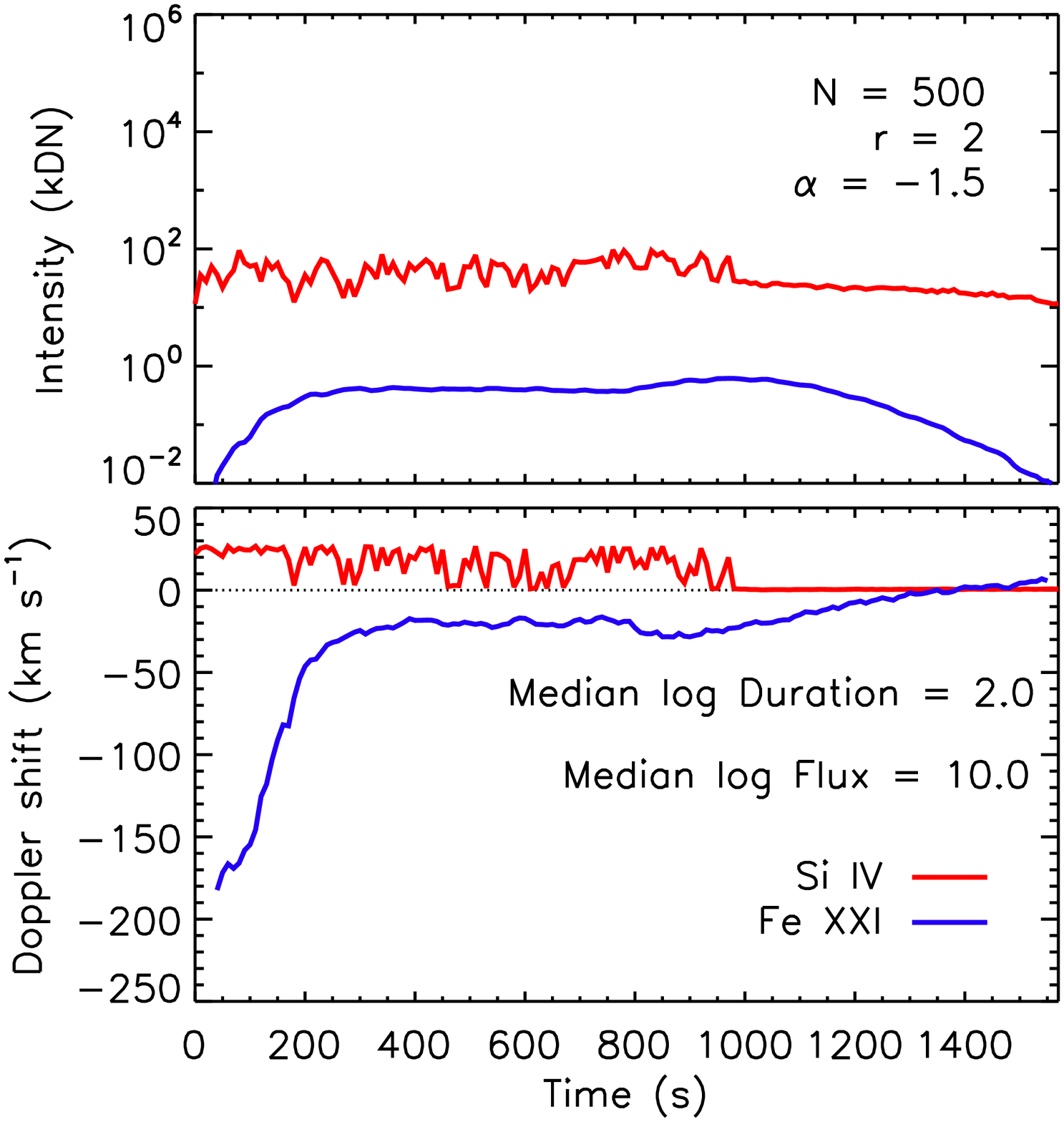}
\end{minipage}
\caption{Similar to Figure \ref{fig:N200}, for 3 cases with 100 seconds of heating and number of loops $N = $[100, 250, 500].  A larger number of loops primarily decreases variability in emission, and produces steadier \ion{Si}{4} red-shifts.  }
\label{fig:N_variable}
\end{figure*}

We briefly examine the effect that the number of loops has on these results.  Choosing the 100-second heating duration case, we vary the number of loops $N$, while holding $N \times r \approx 1000$\,s.  Figure \ref{fig:N_variable} shows 3 such cases, with $N = $[100, 250, 500] loops (and the case with 200 loops is the bottom left plot of Figure \ref{fig:N200}).  In all four cases, the intensities of both lines are similar, though the variability decreases as the number of loops increases.  The Doppler shifts of \ion{Fe}{21} also follow similar trends, with less variability for more loops.  The only major difference is that the red-shifts in \ion{Si}{4} become more pronounced and long-lasting with an increased number of loops, which reiterates a result from \citet{reep2016}.  This is because the newly heated loops at any given time dominate the signal of the line, and these loops are the ones currently experiencing chromospheric condensation.

\subsection{Distribution of Heating Durations}
\label{subsec:distr_dur}

It is probable that the heating durations on all the loops in an arcade are not equal, and therefore there is likely a distribution.  We do not know what distribution that might be, so we therefore briefly examine two cases: a uniform distribution and a power-law distribution, each with an upper and lower limit.  We otherwise follow the same process listed in Table \ref{table:method}, except that in Step 4, we randomly draw a heating duration for each loop instead of using a fixed duration.  

\begin{figure*}
\begin{minipage}[t]{0.32\textwidth}
\includegraphics[width=\linewidth]{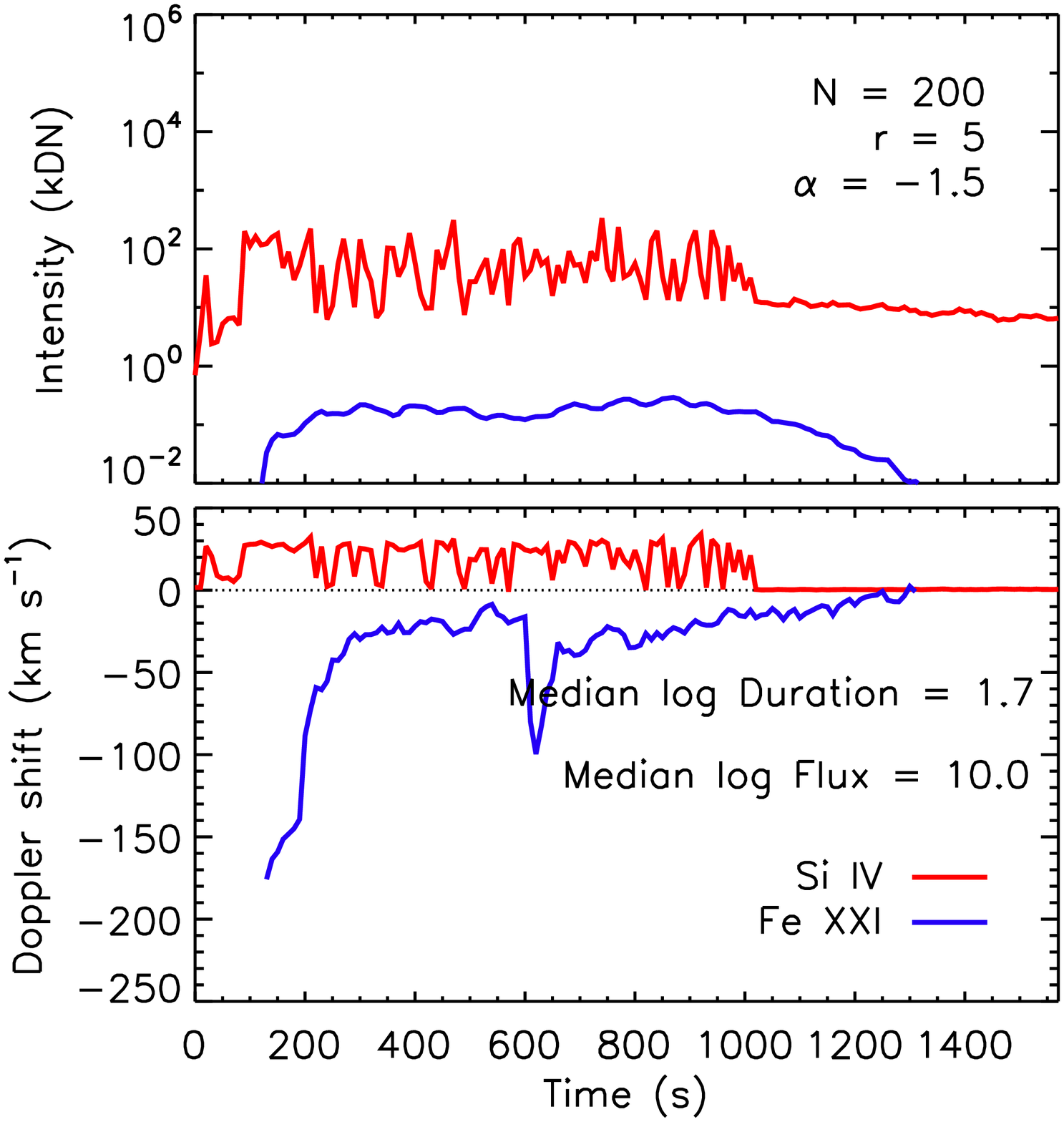}
\end{minipage}
\begin{minipage}[t]{0.32\textwidth}
\includegraphics[width=\linewidth]{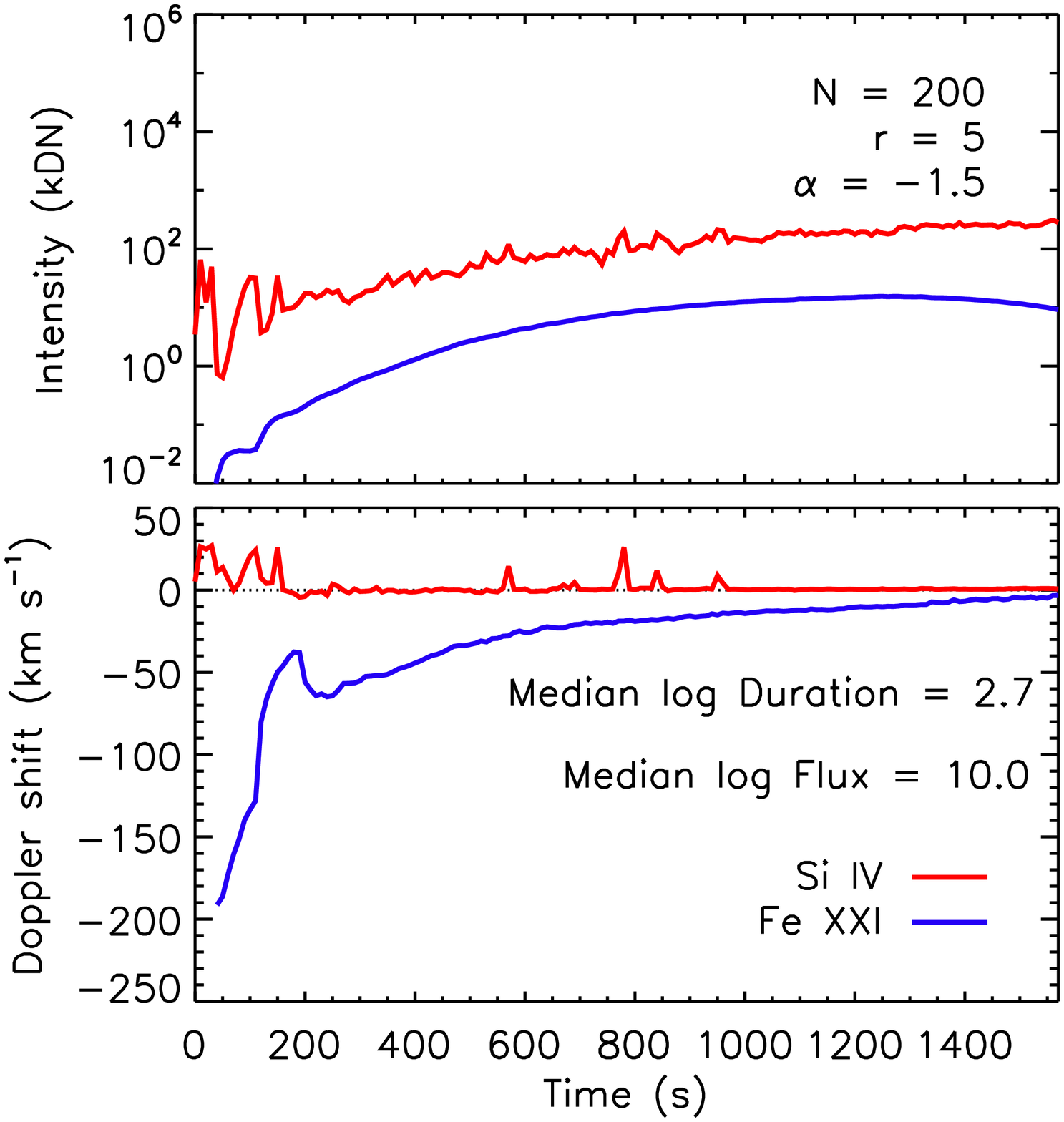}
\end{minipage}
\hspace{\fill}
\begin{minipage}[t]{0.32\textwidth}
\includegraphics[width=\linewidth]{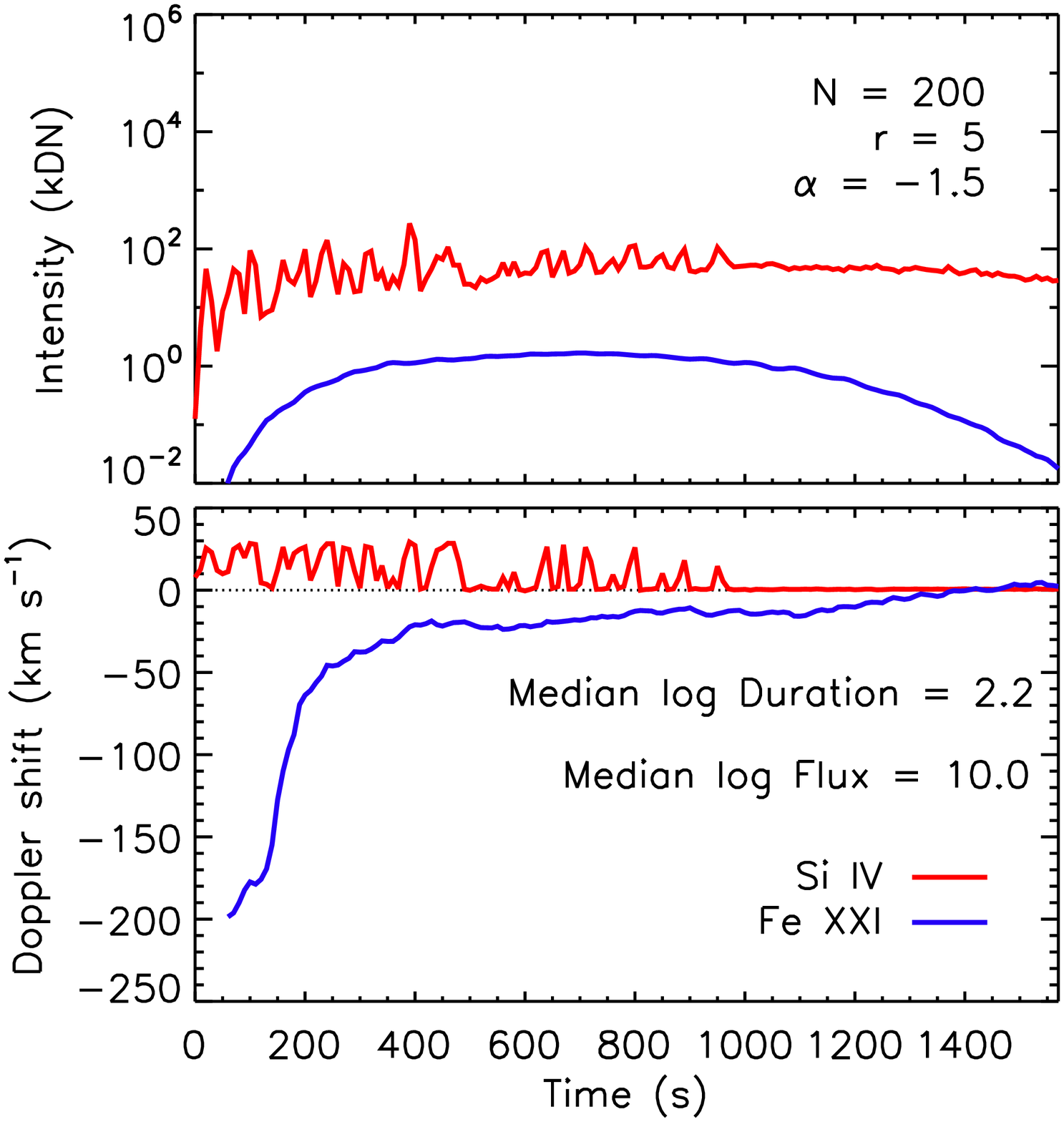}
\end{minipage}
\begin{minipage}[t]{0.32\textwidth}
\includegraphics[width=\linewidth]{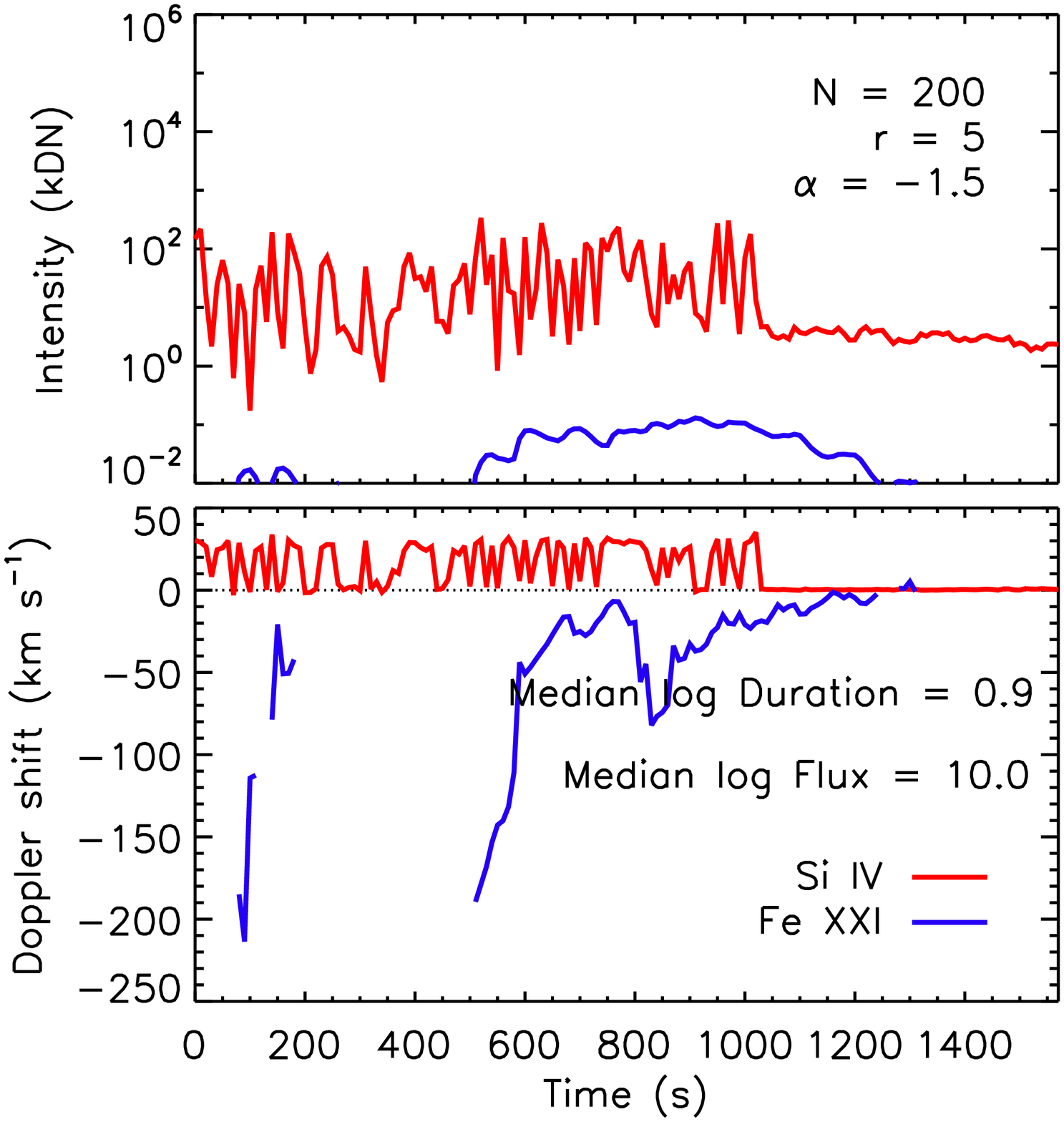}
\end{minipage}
\begin{minipage}[t]{0.32\textwidth}
\includegraphics[width=\linewidth]{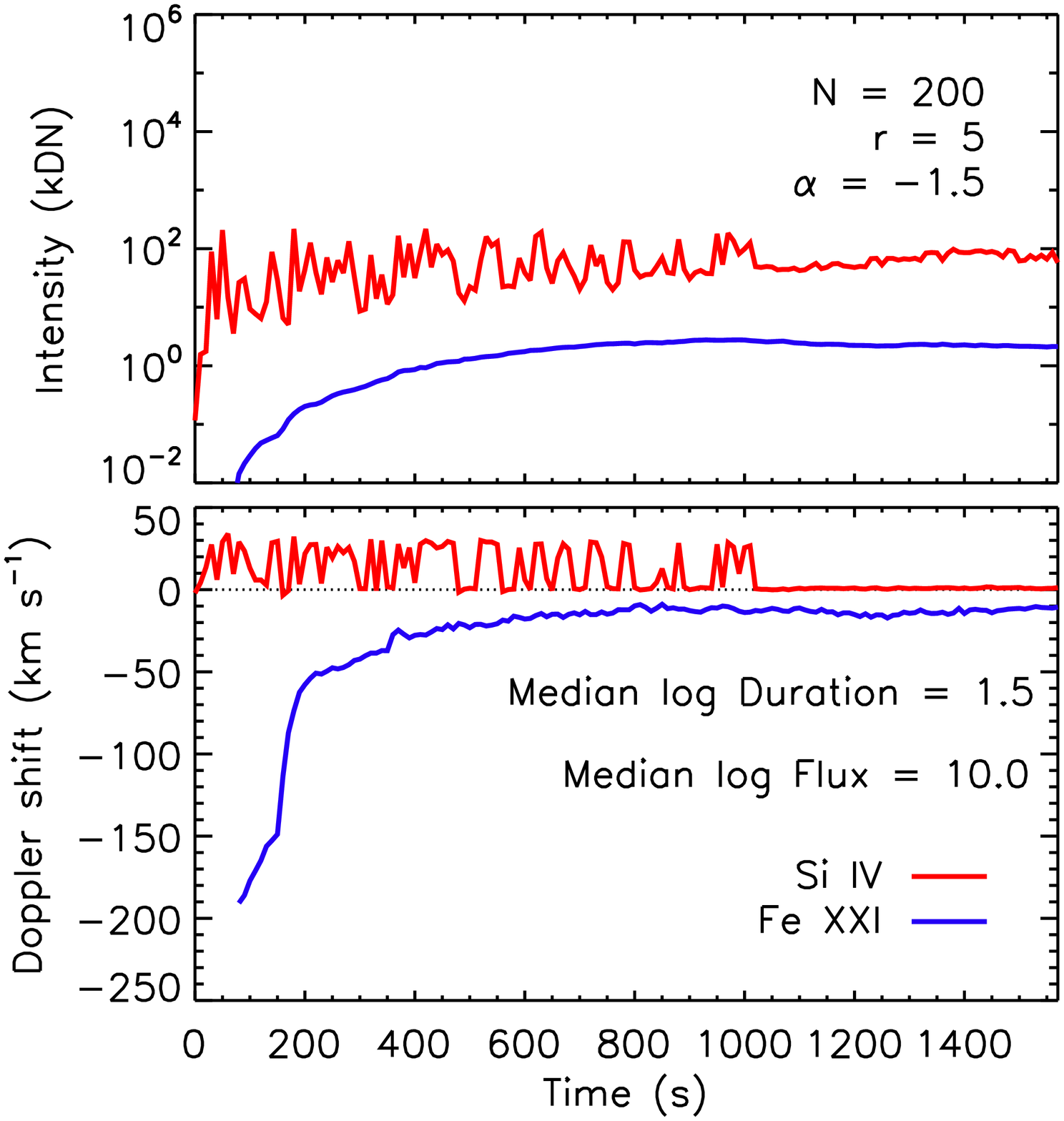}
\end{minipage}
\hspace{\fill}
\begin{minipage}[t]{0.32\textwidth}
\includegraphics[width=\linewidth]{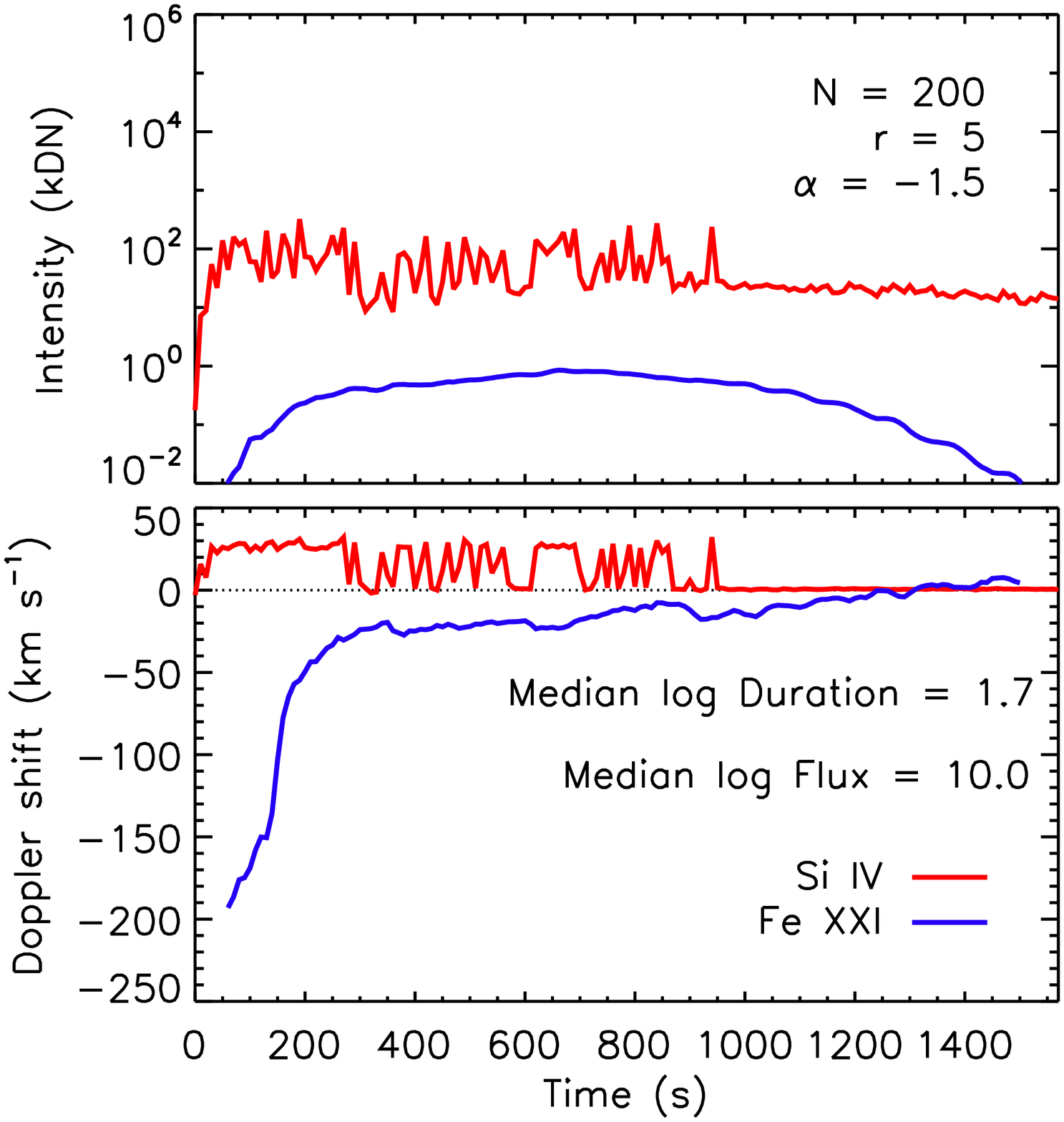}
\end{minipage}
\caption{Synthesized \ion{Si}{4} and \ion{Fe}{21} foot-point emission for 6 multi-threaded simulations with $N = 200$ loops, $r = 5$\,s, $F_{\text{min}} = 3 \times 10^{9}$\,erg\,s$^{-1}$\,cm$^{-2}$, and $\alpha = -1.5$.  The top three plots use a uniform distribution of heating durations, while the bottom three plots use a power-law distribution of slope $-1.0$.  The left column assumes a duration range from 1--100\,s, the middle column 1 -- 1000\,s, and the right column 10--300\,s.  A distribution with a median duration around 50\,s produces reasonable results in both lines.}
\label{fig:distribution}
\end{figure*}

Figure \ref{fig:distribution} shows a few examples.  The top three plots use a uniform distribution of heating durations, while the bottom three use a power-law distribution of heating durations with slope $-1.0$.  The left column assumes a duration range from 1--100\,s, the middle column 1--1000\,s, and the right column 10--300\,s.  

There are two important points to note.  First, as before, extremely long heating durations suppress persistent red-shifts in \ion{Si}{4} (top center), but extremely short heating durations fail to produce \ion{Fe}{21} emission to any appreciable extent (bottom left).  As many flares display both of these signatures (\textit{e.g.} \citealt{sadykov2015,battaglia2015,polito2016}), this suggests that the average heating duration must lie between $\approx 10$--$100$\,s or so.  Second, both distributions produce reasonable results given a reasonable average heating duration and energy flux: compare the top left plot to the bottom right plot, with equal median energy fluxes and median durations of $\approx 50$ and $63$\,s, respectively.  The power-law distribution, however, produces smoother Doppler shifts in \ion{Fe}{21}.  

A large parameter survey could shed more light on the distribution of heating durations.  However, there are many uncertainties and assumptions that limit the usefulness of such an exercise.  We therefore turn our attention to a flare observed with IRIS, and apply this model in order to explain the observed emission.  In this way, we directly check the validity of the model.

\section{\textit{IRIS} observation of the March 12, 2015 flare}
\label{sec:obs}
\begin{figure}
\centering
\includegraphics[width=0.5\textwidth]{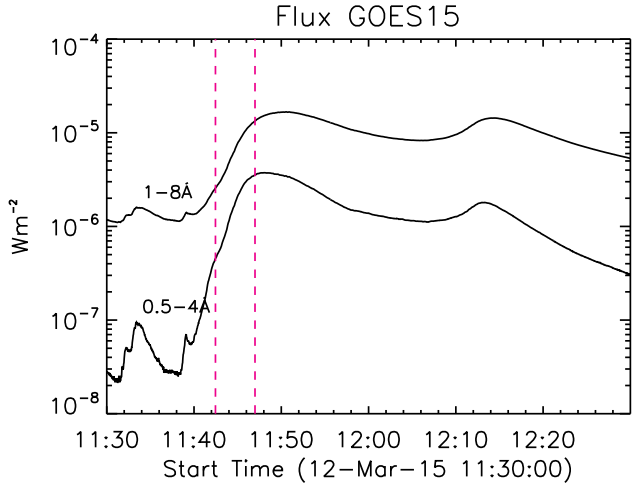}
\caption{Soft X-ray light curves of the M-class flares on March 12, 2015 as observed by the \textit{GOES} satellite in the 0.5--4~\AA~and 1--8~\AA~channels. The dotted pink lines define the time interval over which we measure the blue-shifts of the \ion{Fe}{21} line during the impulsive phase of the first flare.}
\label{fig:goes}
\end{figure}

On March 12, 2015 the \textit{IRIS} spectrograph was observing the AR NOAA 12297 from 05:45 to 17:40~UT during a large sit-and-stare \emph{HOP 245} study with a cadence of about 5~s and exposure time of 4~s.  Two M-class flares occurred between 11:30 and 12:30~UT, peaking at around 11:50~UT (M1.6) and 12:14~UT (M1.4) respectively, as shown in the \textit{GOES} soft X-ray light curves in Figure~\ref{fig:goes}.  These flares were analyzed by several authors \citep[\textit{e.g.}][]{tian2016,brannon2016}.  In this work, we focus on studying the time evolution of the chromospheric evaporation as observed in the \ion{Fe}{21} 1354.08~\AA~line (T~$\approx$~10$^{7}$~K) during the impulsive phase of the first M1.6-class flare, over the interval indicated by the two vertical pink lines in Figure~\ref{fig:goes}.

\textit{IRIS} Slit-Jaw Images (SJI) were taken in three different passbands centered around 1330~\AA, 1400~\AA~and 2832~\AA~with a cadence of about 15~s over a 120\arcsec~x~119\arcsec~field-of-view.  The flare occurred near disk center, at around [-190", -170"].  The three SJI channels are dominated by emission from the \ion{C}{2} 1335.78~\AA~(T~$\approx$~10$^{4.5}$~K), \ion{Si}{4} 1402.77~\AA~(T~$\approx$~10$^{4.9}$~K) and \ion{Mg}{2} wing (T~$\approx$~10$^{3.8}$~K) respectively.  During flares, the SJI 1330~\AA~passband also includes some contribution from \ion{Fe}{21}~1354.08~\AA~emission.  We used level 2 spectral and imaging data, which are processed for dark current subtraction, flat-field and geometry corrections.  The orbital and absolute wavelength calibration of the spectral data was performed by measuring the centroid position of the \ion{O}{1}~1355.568~\AA~photospheric line included in the same spectral window of the \ion{Fe}{21}. 

Images from the SDO/AIA telescope were also analyzed, to provide a context of the observed event and information about the morphology of the flare loops. The AIA level 1 data were processed using the SolarSoft \emph{aia\_prep.pro} routine, which performs the co-alignment of images from different passbands and the adjustment of the telescope plate scale.  The AIA~1600~\AA~images and SJI 1330~\AA~observations, both dominated by chromospheric plasma emission, were co-aligned by eye, giving an uncertainty of around 2 AIA pixels ($\approx$~2\arcsec). 

\begin{figure*}
\centering
\includegraphics[width=\textwidth]{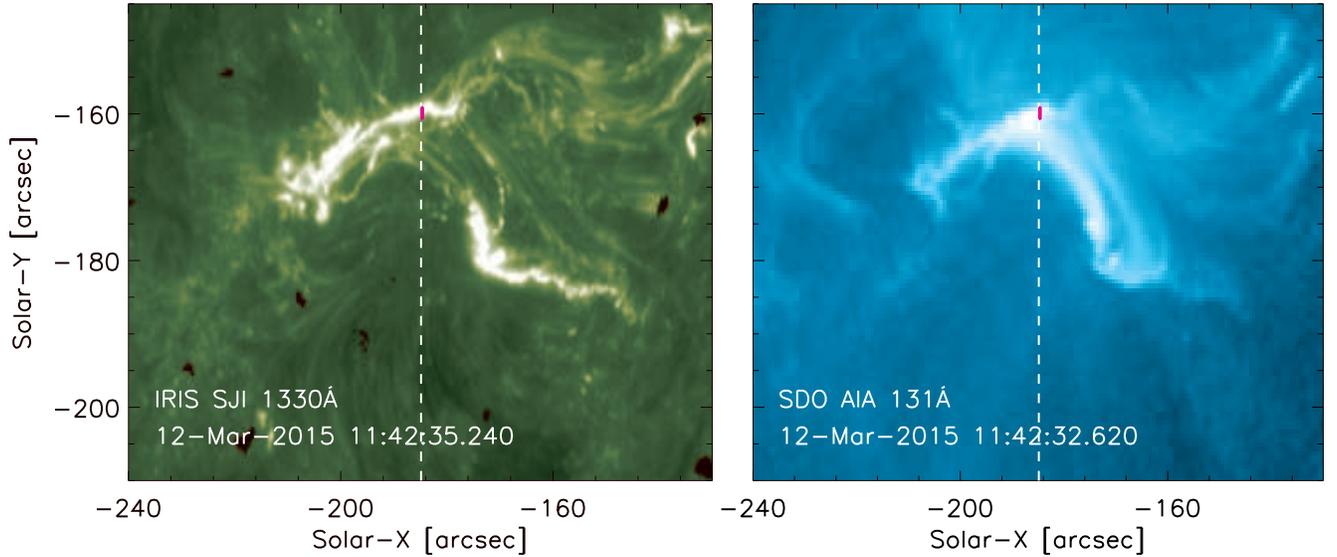}
\caption{\emph{Left panel:} \textit{IRIS} SJI image in the 1330~\AA~channel during the impulsive phase of the flare. The dotted white line indicates the slit position in the IRIS spectrograph sit-and-stare observation. The pink segment highlighted on the slit shows the location on the northern ribbon where we observe the blue-shifts of the \ion{Fe}{21} line over time. \emph{Right panel:} AIA 131~\AA~image closest in time to the SJI image on the left. The location of the \textit{IRIS} slit and the blue-shift location is also overlaid. See Movie 1.}
\label{fig:sji}
\end{figure*}

Figure~\ref{fig:sji} shows the SJI 1330~\AA~image (left) and the closest AIA 131~\AA~image (right) taken at about 11:42:30~UT,  during the early impulsive phase of the M1.6-class flare. The SJI 1330~\AA~filter mainly shows chromospheric emission from the elongated flare ribbons. During flares, the AIA 131~\AA~band is dominated by the hot \ion{Fe}{21} emission from the flare loops.    The vertical dotted line in Fig.~\ref{fig:sji} indicates the IRIS spectrograph slit position of the sit-and-stare study.  It is possible that some \ion{Fe}{21} emission from the hot loops might overlap with the footpoint emission along the line of sight. However, in the early impulsive phase of the flare the spectra at the ribbon are dominated by the blue shifted component due to the evaporation and this component can be easily separated from the almost stationary loop emission.

The evolution of the flare as observed by these two imagers can be best seen in Movie ~1 associated with Fig~\ref{fig:sji}.  As shown in the movie, the IRIS slit crosses the northern flare ribbon and the flare loops during all the observation, and part of the southern ribbon from about 11:58~UT.  Increased emission from the ribbons is observed in the SJI images from $\approx$~11:41:30~UT, whereas the hot (above 10 MK) \ion{Fe}{21} 1354.01~\AA~line can be first clearly detected in the IRIS spectra from about 11:42:30~UT, at the time indicated by the first vertical pink line in Fig.~\ref{fig:goes}. The \ion{Fe}{21} spectra at this time show that the line is very broad ($\geq$~1~\AA) and largely blue-shifted ($\approx$~200 km~$\,$~s$^{-1}$), suggesting ongoing chromospheric evaporation. 

The IRIS spectral data show that the \ion{Fe}{21} emission is then observed to move towards the flare loop top. As a result, the hot emission from the loops becomes progressively more intense, as can be best seen in Movie~1. At the same time, the non-thermal broadening and blue-shift of the line decrease gradually, in agreement with recent IRIS flare observations \citep[e.g.][]{polito2015, polito2016,graham2015}.  This is also consistent with the standard solar flare scenario, suggesting that the flare loops are filled with evaporating plasma from the flare foot-points and ribbons. 

In Section \ref{subsec:obs_bs} we will analyze the time evolution of the evaporating hot plasma from the flare ribbons, as observed in the \ion{Fe}{21} 1354.08~\AA~spectra. The results will then be compared to the predictions of our flare simulations in Section \ref{subsec:comparison}.

\subsection{Evolution of \ion{Si}{4} and \ion{Fe}{21} shifts}
\label{subsec:obs_bs}
\begin{figure*}
\centering
\includegraphics[width=0.4\textwidth]{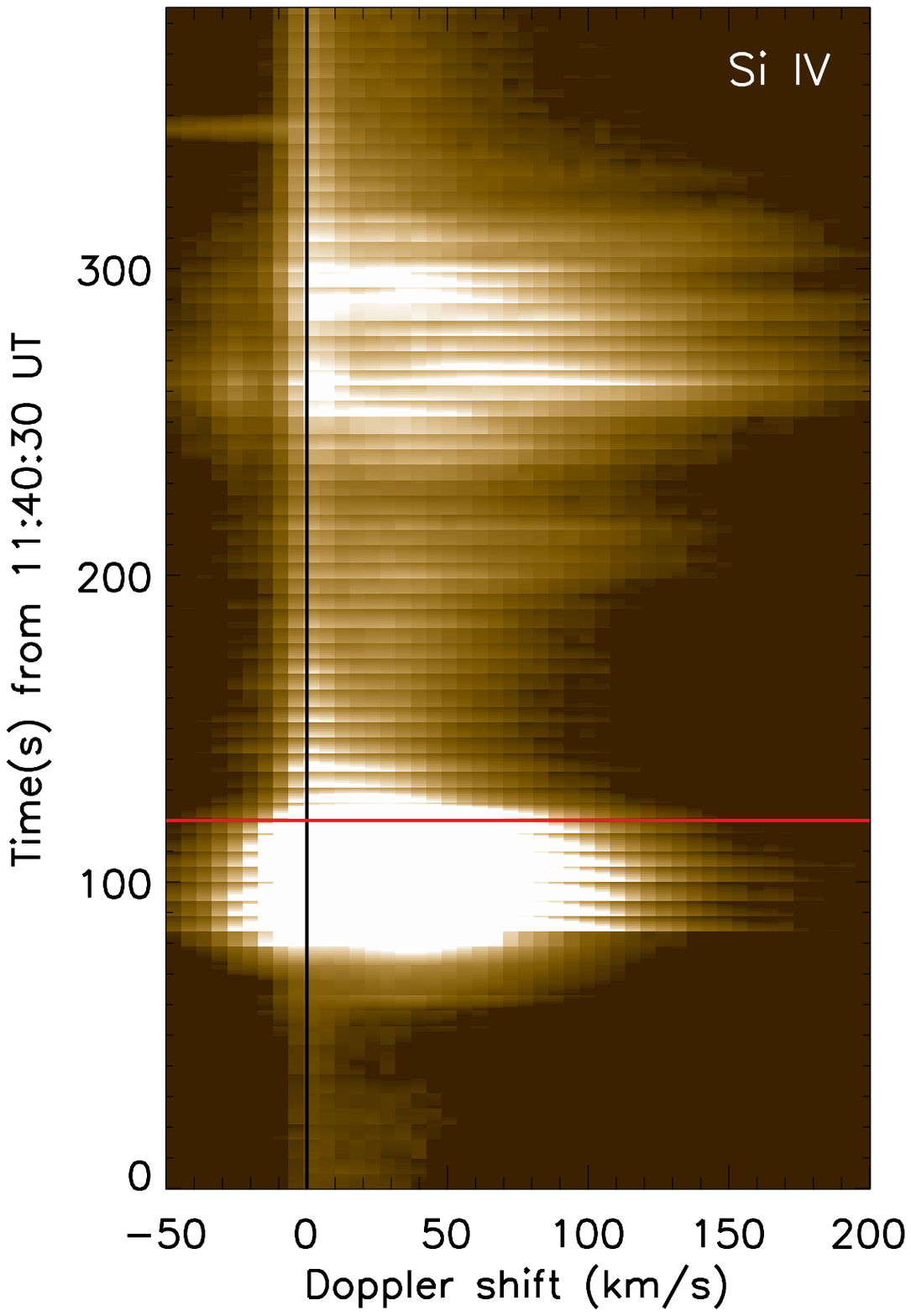}
\includegraphics[width=0.4\textwidth]{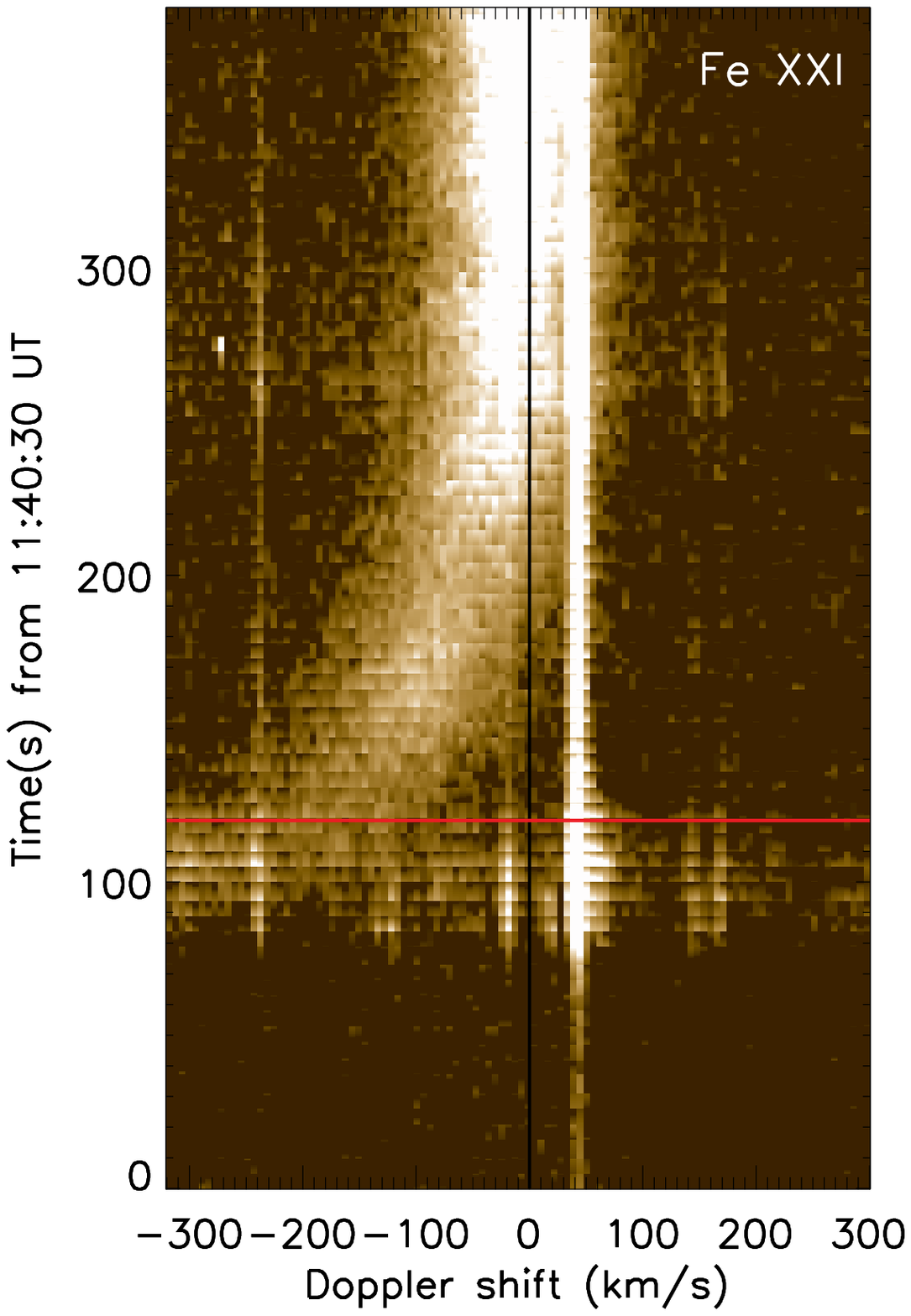}
\caption{Time evolution of the \ion{Si}{4} and \ion{Fe}{21} spectral lines as observed just above the ribbon location for the March 12, 2015 flare. The plots are obtained by stacking together a slice (around the slit-pixels 284 and 285) of the CCD images of the spectral windows as a function of time. The spectra are plotted as a function of Doppler shift velocity, where negative (positive) values indicate blue-shifts (red-shifts) of the line from the rest wavelengths of 1402.77~\AA and 1354.01~\AA.  The red lines indicate the start time for Figure \ref{fig:obs_doppler}.  }
\label{fig:obs_spectra}
\end{figure*}

\begin{figure}
\centering
\includegraphics[width=0.48\textwidth]{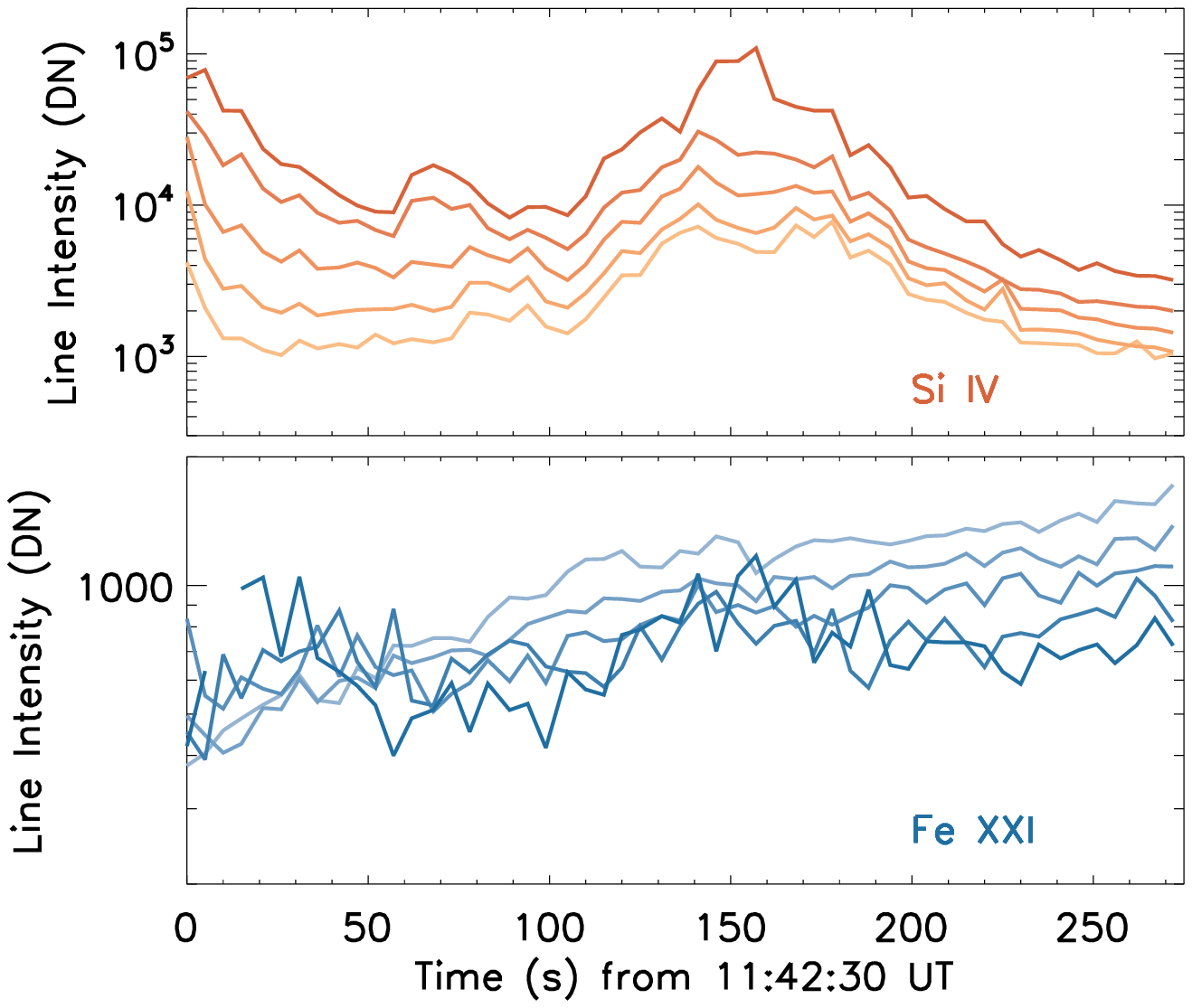}
\includegraphics[width=0.48\textwidth]{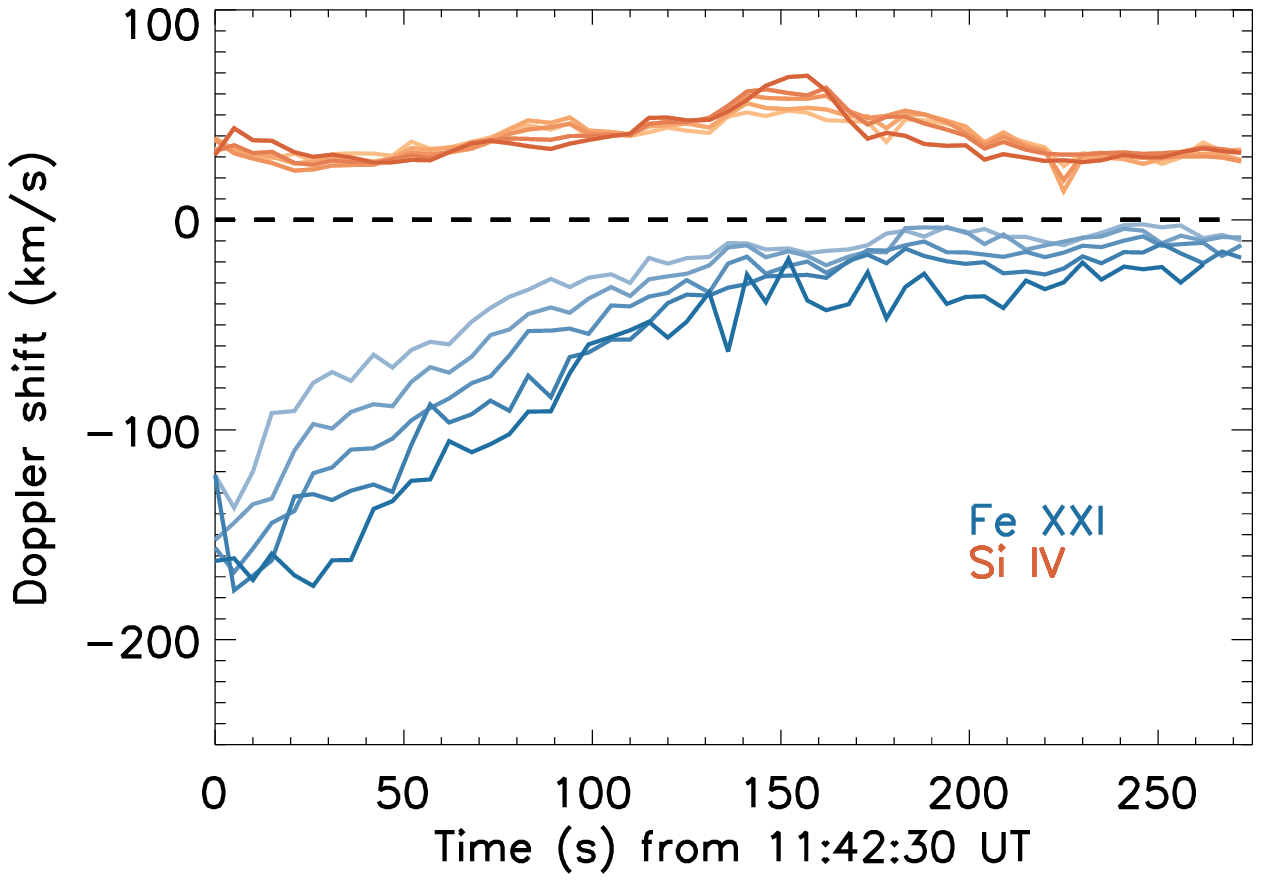}
\caption{Intensity (top panel) and velocity (bottom panel) of the \ion{Si}{4} and \ion{Fe}{21} lines as a function of time for 5 pixels close to the location of the flare ribbon.  Different colors indicate different pixels.  Negative values of Doppler shift indicate blue-shifts.  The start time is shown as a red line in Figure \ref{fig:obs_spectra}.  }
\label{fig:obs_doppler}
\end{figure}

The maximum blue-shift of the \ion{Fe}{21} is observed just above the intense FUV continuum emission from the flare ribbons, between the IRIS slit pixels 283 and 287. These slit pixels correspond to the location on the Sun which is highlighted in pink in Fig.~\ref{fig:sji}.  The \ion{Si}{4} line is not saturated in these pixels, in contrast to the intense ribbon location.  Fig.~\ref{fig:obs_spectra} shows the time evolution of the \ion{Si}{4}(left) and \ion{Fe}{21} (right) spectra in that location, as obtained by stacking together slices of the CCD images over time. We performed a fit with a single Gaussian of the \ion{Si}{4} and \ion{Fe}{21} spectral window for each of these 5 slit pixels over time using the Solarsoft routine \emph{xcfit\_block.pro}. Figure~\ref{fig:obs_doppler} shows the intensity (top) and Doppler shift centroid velocity (bottom) of the lines as a function of time, as obtained by the fitting procedure. Different colors of the plot symbols represent the results for different IRIS slit pixels, from 283 to 287. Negative (positive) values of Doppler velocity indicate blue-(red-)shifts of the \ion{Si}{4} and \ion{Fe}{21} lines from their at-rest observed wavelength of $\approx$~1402.77 and 1354.1~\AA~respectively. The reference wavelength of the \ion{Fe}{21} was measured during the gradual phase of the flare at the loop top, where the line is expected to be at rest.  The red lines indicate the start time for Figure \ref{fig:obs_doppler}.

Figure~\ref{fig:obs_doppler} shows that the blue-shift of the \ion{Fe}{21} line during the M-class flare under study gradually decreases (while its intensity increases) going towards the peak of the flare. The line becomes completely stationary in most of the pixels in about 4~min or more, at the time indicated by the second pink line in Fig.~\ref{fig:goes}.  Previous work by \cite{graham2015} and \cite{polito2016} found an evaporation duration of $\approx$~10min for two different X-class flares \citep[see also e.g.][]{li2015}. 

\subsection{Comparison to the multi-threaded model}
\label{subsec:comparison}

\begin{figure*}
\centering
\includegraphics[width=0.48\textwidth]{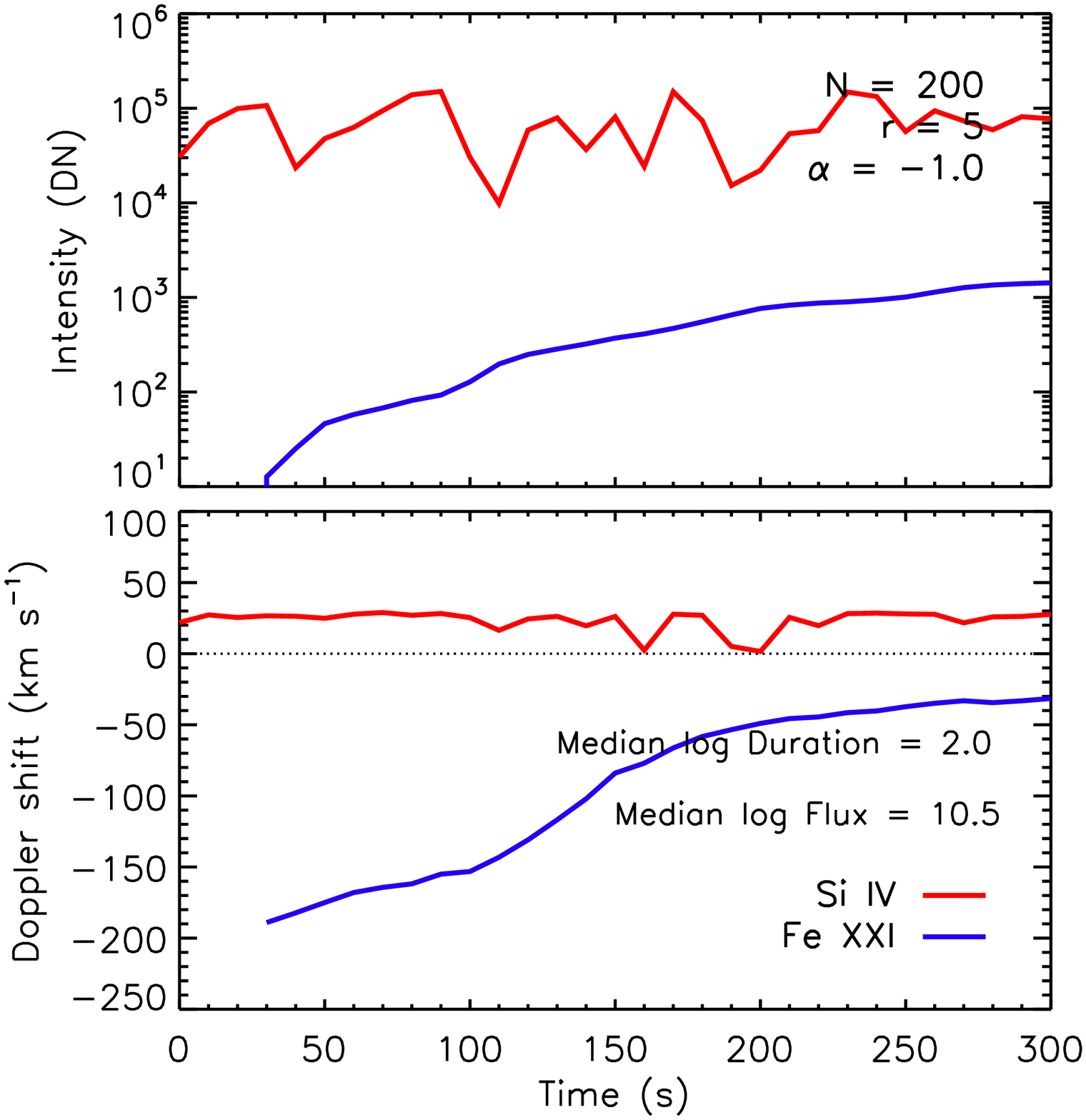}
\includegraphics[width=0.48\textwidth]{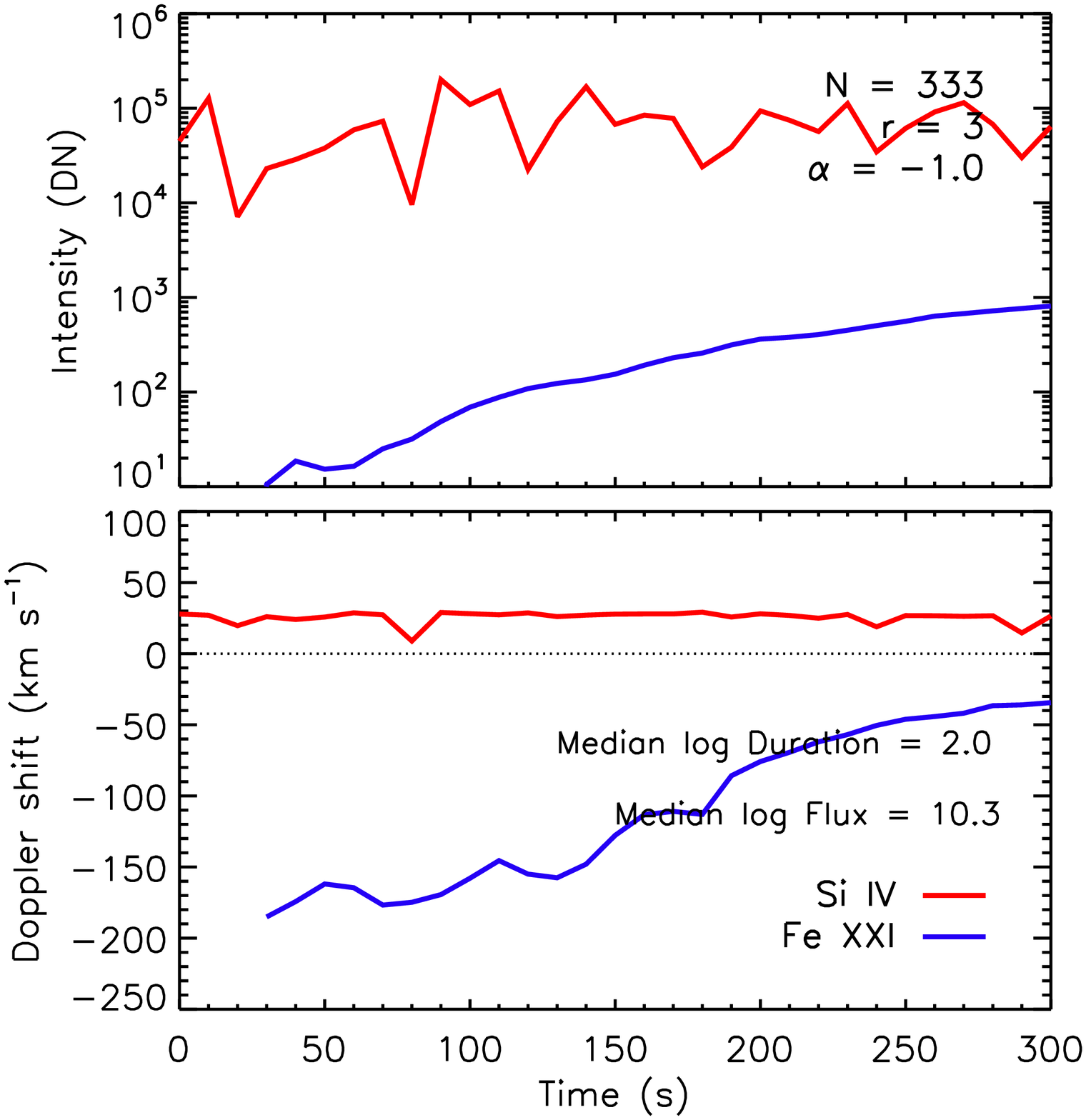}
\caption{The modeled foot-point emission from two sets of multi-threaded simulations with $r = 5$ (left) and $3$\,s (right), with a high median flux.  Compare the intensities and Doppler shifts to the observed quantities in Figure \ref{fig:obs_doppler}, with which there is a broad consistency.   }
\label{fig:model_doppler}
\end{figure*}

We briefly compare the observations against the multi-threaded model.  Using a power-law distribution of heating durations between 30 and 300\,s, we model two cases: $r = 5$ and $3$\,s, shown in Figure \ref{fig:model_doppler}.  The modeled values are mostly consistent with the observations, with some differences.  The intensities in both \ion{Si}{4} and \ion{Fe}{21} are approximately the same ($10^{5}$ and $10^{3}$\,DN at their peaks, respectively).  The observed \ion{Si}{4} intensities appear to have a decreasing trend, however, suggesting that the energy input is decreasing with time, which is not accounted for in the simulations.  The red-shifts in \ion{Si}{4} remain close to $30$\,km\,s$^{-1}$ for the time period under consideration.  The \ion{Fe}{21} intensities grow with time, while the blue-shifts gradually decay over about 4--5\,min from a maximum of about $200$\,km\,s$^{-1}$.  Finally, the initial observed intensities of \ion{Fe}{21} are higher and have a slightly more gradual change with time than the simulations.  Overall, the basic premise of the model is consistent with the observations, although the match is far from perfect due to our lack of knowledge of the exact parameters on the sun.

\section{Discussion}
\label{sec:discussion}

In this work, we have shown that the heating duration plays a vital role on the observed Doppler shifts and intensities of the \ion{Fe}{21} and \ion{Si}{4} lines observed routinely by \textit{IRIS} in solar flares.  In order to produce decays of 5--10 minutes of blue-shifts in \ion{Fe}{21} as in \citet{graham2015}, it is necessary to heat a loop for a similar duration.  If a loop were heated for a short time with a large enough energy flux, it is possible to produce \ion{Fe}{21} emission, but the evaporative up-flows only last slightly longer than that heating duration.  The logic is simple: once the heating ceases, the over-pressure that causes the expansion of material also ceases.  We therefore find that longer heating durations are required to explain the observations (within the multi-threaded modeling we use), and suggest that the duration of up-flows act as a diagnostic of that heating duration.  In earlier work, \citet{warren2006} came to a similar conclusion: heating durations on individual threads of $\approx 200$\,s are more consistent with GOES and Yohkoh soft X-ray light curves than heating durations of $\approx 20$\,s.

For loops heated strongly enough to produce strong evaporative up-flows, there must also be down-flows due to the conservation of momentum.  Therefore, we simultaneously expect to measure red-shifts in cooler lines like \ion{Si}{4}.  Indeed, we do find such red-shifts, though they are significantly longer lived than the roughly 60\,s predicted by \citet{fisher1989}.  A multi-threaded model, with multiple loops rooted in one pixel being heated in succession, naturally explains these pervasive red-shifts as the succession of chromospheric condensations on each successive loop.  This was the primary result of \citet{reep2016}.  When both \ion{Si}{4} and \ion{Fe}{21} emission are present in the data, we can constrain the number of loops within a pixel, the energy fluxes onto those loops, and the heating duration on those loops.

This method can be generalized for a detector with a wider temperature coverage and similar cadence.  For example, \ion{Fe}{23} up-flows behave similarly in some flares (\textit{e.g.} \citealt{brosius2013}), where the higher temperature of formation for that line could be used to more strongly place limits on the energy flux.  Further, the amount of plasma at temperatures exceeding 20--30\,MK requires large energy fluxes, so that the relative proportion of this super-hot component could act as another test for this model.  Constraining the energy flux in this way is an important diagnostic, as there are no other direct ways to measure this value on individual loops, and the evolution of plasma on a loop is primarily determined by the energy input.
%
%

We therefore summarize the results:
\begin{enumerate}
\item[(1)] The duration of chromospheric evaporation depends intimately on the heating duration.  To produce evaporation lasting 5 -- 10\,min requires heating durations nearly as long.
\item[(2)] There is a distribution of heating durations, with average values $\approx 50$ -- $100$\,s consistent with the data.  If the average value is too long, persistent red-shifts seen in \ion{Si}{4} are suppressed.  If the average value is too short, \ion{Fe}{21} emission does not form in appreciable amounts, and the evaporation decays too quickly.
\end{enumerate}

\leavevmode \newline

\acknowledgments  This research was performed while JWR held an NRC Research Associateship award at the US Naval Research Laboratory with the support of NASA.  V.P. was supported by NASA grant NNX15AF50G, and by contract 8100002705 from Lockheed-Martin to SAO.  Figure \ref{fig:apex_values} was produced with color-blind friendly IDL color tables kindly provided by Graham Kerr and Paul Wright (see \citealt{pjwright}).  This research has made use of NASA's Astrophysics Data System.  CHIANTI is a collaborative project involving George Mason University, the University of Michigan (USA) and the University of Cambridge (UK).  IRIS is a NASA small explorer mission developed and operated by LMSAL with mission operations executed at NASA Ames Research center and major contributions to downlink communications funded by ESA and the Norwegian Space Centre.
\bibliography{apj}
\bibliographystyle{aasjournal}

\end{document}